\pdfoutput=1
\documentclass[aip,amsmath,amssymb,reprint]{revtex4-1}

\usepackage{graphicx}% Include figure files
\usepackage{dcolumn}% Align table columns on decimal point
\usepackage{bm}% bold math
\usepackage[utf8]{inputenc}
\usepackage[T1]{fontenc}
\usepackage{mathptmx}

\begin{document}

\preprint{AIP/123-QED}
\title[]{Effect of Topology upon Relay Synchronization in Triplex Neuronal Networks}
%\title[Relay Synchronization of Chimera States]{Relay Synchronization of Chimera States in Neuronal Triplex Networks}
% Force line breaks with \\

\author{Fenja Drauschke}
\author{Jakub Sawicki}%
\affiliation{ 
Institut f\"ur Theoretische Physik, Technische Universit\"at Berlin, Hardenbergstr, 36, 10623 Berlin, Germany
}%

\author{Rico Berner}
\email{rico.berner@physik.tu-berlin.de}
\affiliation{ Institut f\"ur Theoretische Physik, Technische Universit\"at Berlin, Hardenbergstr, 36, 10623 Berlin, Germany
}%
\affiliation{Institut f\"ur Mathematik, Technische Universit\"at Berlin, Strasse des 17. Juni 136, 10623 Berlin, Germany.}

\author{Iryna Omelchenko}

\author{Eckehard Sch\"oll}
\email{schoell@physik.tu-berlin.de}
\affiliation{ 
Institut f\"ur Theoretische Physik, Technische Universit\"at Berlin, Hardenbergstr, 36, 10623 Berlin, Germany
}

\date{\today}% It is always \today, today,

\begin{abstract} %limit 250 words
\section*{}%Problemstellung

Relay synchronization in complex networks is characterized by the synchronization of remote parts of the network due to their interaction via a relay. In multilayer networks, distant layers that are not connected directly can synchronize due to signal propagation via relay layers. In this work, we investigate relay synchronization of partial synchronization patterns like chimera states in three-layer networks of interacting FitzHugh-Nagumo oscillators. We demonstrate that the phenomenon of relay synchronization is robust to topological random inhomogeneities of small-world type in the layer networks. We show that including randomness in the connectivity structure either of the remote network layers, or of the relay layer, increases the range of interlayer coupling strength where relay synchronization can be observed.
\end{abstract}

\maketitle

\begin{quotation}

The investigation of synchronization in networks of coupled oscillatory units is a vivid research area with broad applications in nature and technology~\cite{PIK01,BOC14}. Uncovering complex mechanisms leading to coexistent synchrony and asynchrony plays an important role in our understanding of many biological~\cite{GOL11a} and technological systems~\cite{FIS06,BER12}. Collective phenomena such as remote synchronization where distant parts of complex networks synchronize despite of the absence of a direct connection is a challenging problem. Particularly, in multilayer networks, remote layers can synchronize due to their interaction through intermediate (relay) layers. Such scenarios have been observed in the setting of multiplex networks~\cite{LEY18}. These are special multilayer networks where each layer contains the identical set of nodes, and only one-to-one connections between the corresponding nodes in the neighboring layers are allowed. Depending on the type of network nodes and intralayer topologies, complex partial synchronization patterns such as clusters, chimera states, or solitary states can be formed inside the layers. Only recently, relay synchronization of chimera states, i.e., states with spatially coexisting coherent and incoherent domains, has been explored and shown to be of interest for a wide range of application~\cite{SAW18c}. In the present work, we analyze the robustness of the relay synchronization of chimera states in three-layer multiplex networks of FitzHugh-Nagumo oscillators, which are a paradigmatic model widely used in neuroscience. 
We find that relay synchronization is sensitive to changes in the connectivity of the remote layers as well as the relay layer.

\end{quotation}

%\textbf{***RB:I would take out the word regular everywhere, because all networks considered here are regular form the graph theoretical point of view. We can use ring, ring structure, nonlocla structure, etc. instead*** FD: I have mostly replaced regular, but sometimes left 'regular ring' you can decide if this is alright.}
\section{Introduction}

The analysis of complex networks is of great interest with respect to various real-world systems such as social networks \cite{GIR02}, economics \cite{MOS16}, ecology \cite{GRI05}, finance \cite{SPR04a}, transport systems \cite{WOO11, CAR13d} as well as the neural activity in the brain \cite{BEN16,BAT17,CHO18,RAM19}. Many of such systems can be represented as multilayer networks~\cite{KIV14,DOM16a}, where elements are organized in layers with different types of interaction within and between the layers. Multiplex networks represent a special type of multilayer structures, where each layer contains the same number of nodes and only one-to-one connections between equivalent nodes from neighboring layers are permitted. Multiplex networks can be associated with social and technological systems~\cite{BOC18}, for instance with dynamical processes where layers correspond to the states of the system at different times~\cite{MUC10}, or with different types of links. Multilayer structures allow for full~\cite{TAN19} or partial~\cite{BER19b} synchronization between the layers, and different synchronization scenarios can be observed, such as explosive synchronization~\cite{ZHA15a}.

An intriguing phenomenon in networks with multiplex topology is relay (or remote) synchronization between layers that are not directly connected and interact via an intermediate (relay) layer. This can occur in single systems, such as lasers \cite{FIS06} and electronic circuits \cite{BER12}, as well as in networks between remote pairs of nodes~\cite{NIC13, GAM13} or between remote pairs of layers~\cite{ZHA17,SAW18c, LEY18,SAW19a, WIN19}. Relay synchronization allows for distant coordination, which may be applied to encryption key distribution and secure communication \cite{ZHA17}. Beside this, it is of great relevance in human brain networks, where the thalamus \cite{GUI02}, as well as the hippocampus \cite{GOL11a} are known to act as a relay between different brain areas. While previous work on relay synchronization used simple nonlocally coupled ring networks \cite{SAW18c}, the human brain is organized by small-world topologies on the macro- as well as on the micro-scale~\cite{BAS06,BAS06a}. Therefore, in this work we generalize relay synchronization to more complex, small-world-like topologies. Thereby, we also gain a deeper understanding of the underlying mechanisms of relay synchronization, which could give insight into the functionality of certain brain processes.

With regard to relay synchronization in complex networks, Ref.~\onlinecite{GAM16b} has revealed the significance of network structural and dynamical symmetries for the appearance of distant synchronization. However, while some dynamical aspects of remote synchronization of individual nodes have been uncovered~\cite{NIC13, PEC14, LEY18}, very little is known about relay synchronization of partial synchronization patterns. Prominent examples of such patterns are chimera states, characterized by the coexistence of synchronized (coherent) and desynchronized (incoherent) spatial domains. Chimera states were found in a plethora of networks of coupled oscillators \cite{KUR02a,ABR04,PAN15,SCH16b,OME18a,OME19c,SCH20b}. The coexistence of synchrony and asynchrony in the brain is reminiscent of the phenomenon of unihemispheric sleep\cite{RAT00,RAT16,MAS16,RAM19}. Similarly, during perceptual organization \cite{NIK10a}, behavioral sensation \cite{AHN13, AHN14}, and epileptic seizures \cite{JIR13,JIR14,CHO18} partially synchronized patterns arise, which can be associated with chimera states. Chimera patterns have been numerically observed in networks of coupled neurons with nonlocal topologies \cite{SAK06a,OME13,HIZ13,OME15,HIZ16}.

The purpose of this work is to uncover the effects of the layer topologies upon the scenarios of relay synchronization in a three-layer network consisting of FitzHugh-Nagumo oscillators \cite{FIT61}. Here, the individual layers are initially organized as nonlocally coupled rings allowing for the formation of chimera states within the layers~\cite{OME13,OME15}. In such networks, two remote layers synchronize due to their interaction via the relay layer. Moreover, a special regime of "double chimera" is observed where only coherent domains of the chimera states synchronize remotely~\cite{SAW18c}. We examine the robustness of relay synchronization by perturbing the network topologies within the layers. Applying the Watts-Strogatz algorithm~\cite{WAT98}, first, we randomly remove links in the remote layers and replace them with random shortcuts. Second, we consider topological inhomogeneity in the relay layer keeping the remote layers as regular rings. Surprisingly, we observe that introducing topological inhomogeneities increases the range of the interlayer coupling strength for which relay synchronization in the network takes place.

\section{The Model}
\begin{figure} %[h!]
    \centering
    \includegraphics[width=6cm]{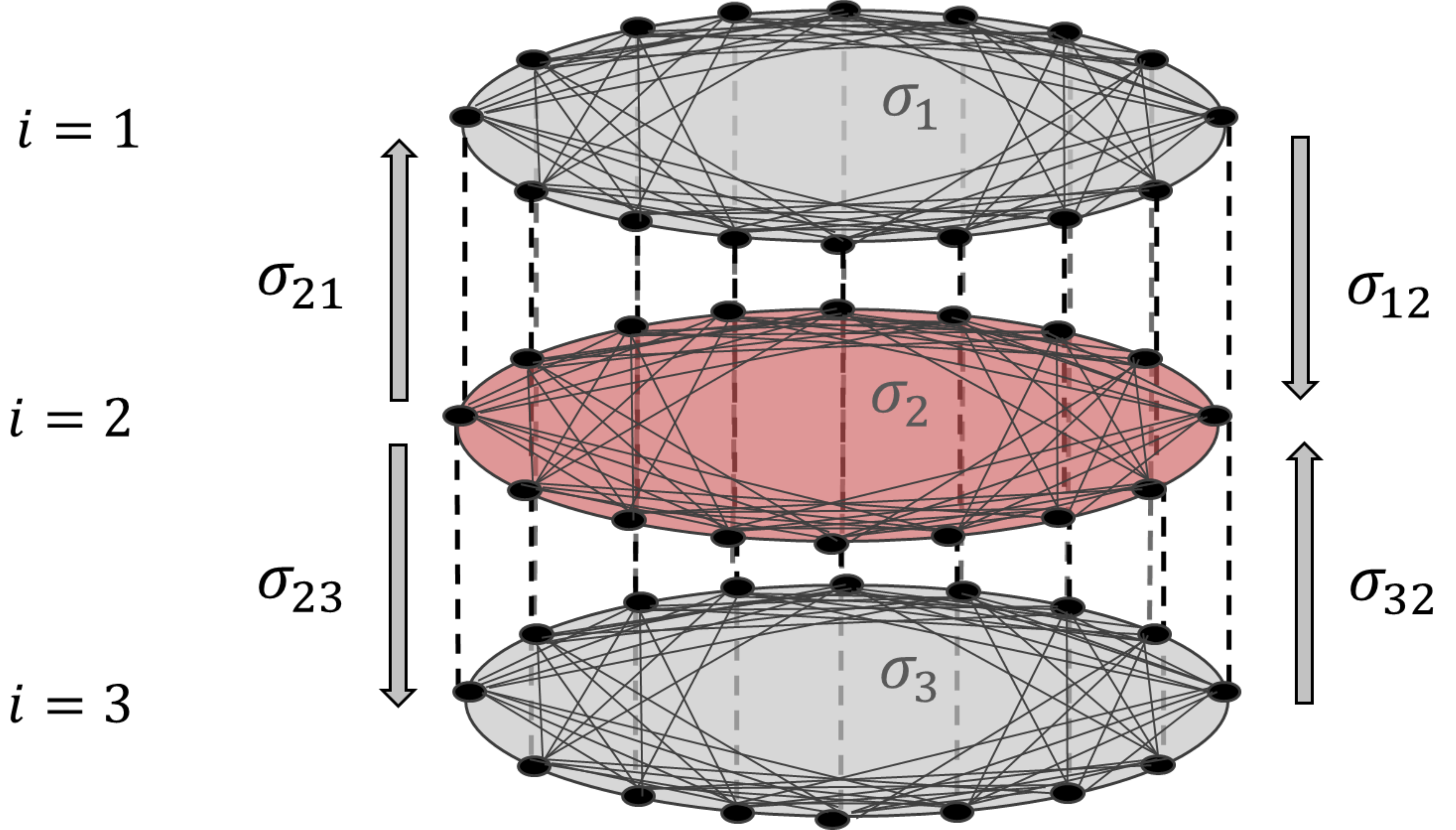}
    \caption{Illustration of a triplex network with nonlocally coupled ring topology with intralayer coupling strength $\sigma_i$ and interlayer coupling strength $\sigma_{ij}$ ($i,j=1,2,3$). The remote layers ($i=1,3$), depicted in grey are connected through the relay layer ($i=2$), marked in red.
    }
    \label{network_model}
\end{figure}

We consider a multiplex network consisting of three layers, schematically shown in Fig.~\ref{network_model}. Each layer, labeled by $i=1,2,3$, contains $N$ nodes which form a nonlocally coupled ring where each node is connected with its $R$ nearest neighbors to both sides. Throughout the paper we fix $N=500$ and $R=170$. Between the layers, the nodes are bidirectionally connected with their corresponding counterpart from the neighboring layers.

The dynamical variable $\mathbf{x}_k^i= (u_k^i, v_k^i)^{T}$ of each node is governed by
\begin{align}
\dot{\mathbf{x}}_k^i(t)=\mathbf{F}(\mathbf{x}_k^i(t))+\frac{\sigma_i}{R}\sum_{l=1}^{N}a^i_{kl}\mathbf{H}[\mathbf{x}^i_k(t)-\mathbf{x}^i_l(t)]\nonumber\\+\sum_{j=1}^{3} \sigma_{ij}\mathbf{H}[\mathbf{x}^j_k(t)-\mathbf{x}^i_k(t)],
\label{dynamics_eq_end}
\end{align}
where $k=1,...,N$ numbers the nodes, $i=1,2,3$ labels the layer, and $a^i_{kl}\in[0,1]$ are the elements of the adjacency matrix $\mathbf{A}^i$, determining the fundamental topology of layer $i$, $\sigma_i$ is the intralayer coupling strength, and $\sigma_{ij}$ is the interlayer coupling strength. The local dynamics of each oscillator is governed by the FitzHugh-Nagumo (FHN) system: 
\begin{align}
\mathbf{F}(\mathbf{x})=\binom{ \frac{1}{\epsilon} (u- \frac{u^3}{3}-v)}{u+a} 
\label{FHN_uv_eq}
\end{align}
where $u$ is the activator (membrane potential), and $v$ is the inhibitor (recovery variable comprising all inhibitory processes). We fix the threshold parameter $a=0.5$ and the time-scale separation parameter $\epsilon =0.01$ throughout the paper, ensuring oscillatory dynamics of the individual units. The interlayer coupling scheme for the triplex network is defined as follows:
\begin{align}
\boldsymbol{\sigma}=
    \begin{pmatrix}
        0 & \sigma_{12} & 0 \\
        \sigma_{21} & 0 & \sigma_{23} \\
        0 & \sigma_{32} & 0 \\
    \end{pmatrix}.
    \label{sigma_matrix}
\end{align}
By choosing $\sigma_{12}$ = $\sigma_{32}$ and $\sigma_{21}$= $\sigma_{23}$= $\frac{\sigma_{12}}{2}$ the interlayer coupling is bidirectional and has a constant row sum. Throughout this paper the interaction between the neurons is realized through a rotational coupling matrix in order to include also  cross-coupling between activator and inhibitor.
\begin{align}
%\mathbf{H}_{ij}:=
\mathbf{H}=
\begin{pmatrix}
    \epsilon^{-1}\cos\phi & \epsilon^{-1}\sin\phi\\
    -\sin\phi & \cos\phi
\end{pmatrix}
\label{H_matrix}
\end{align}
Fixing the coupling phase to $\phi = \frac{\pi}{2} -0.1$ allows for the observation of chimera states within the layers \cite{OME13}.

%\section{Measures}
%\subsection{Synchronization}
To analyze and distinguish different forms of synchronization between the layers in our networks, we employ the following measures. The global interlayer synchronization error $E^{ij}$ quantifies the synchronization between layers $i$ and $j$:
\begin{equation}
    E^{ij}= \lim_{T \to \infty}\frac{1}{NT}\int_0^T \! \sum_{k=1}^{N}  \, \left\lVert \mathbf{x}_k^i(t)-\mathbf{x}_k^j(t)\right\rVert \mathrm{d}t,
    \label{E_glob_eq}
\end{equation}
where $\Vert \cdot \Vert$ is the Euclidean norm, and $i,j=1,2,3$ denote to the layer number\cite{SAW18c}. The global interlayer synchronization error is equal to zero when all corresponding oscillators in two layers are synchronized and perform identical dynamics. Based on this measure, one can distinguish the following types of synchronization:
\begin{enumerate}
\item\textbf{Full interlayer synchronization:}  
All layers are completely synchronized ($E^{12}=E^{13}=0$).
\item\textbf{Relay interlayer synchronization:} 
Remote layers are completely synchronized, and not synchronized to the relay layer ($E^{13}=0$ and $E^{12}\neq 0$).
    \item\textbf{Interlayer desynchronization:}
Different patterns in each layer ($E^{12} \neq E^{13} \neq 0$)
\end{enumerate}

While the global interlayer synchronization error gives us information about the complete synchrony of the layers, it is not able to distinguish possible regimes of partial synchronization between the layers. For the purpose of obtaining more detailed information, we analyze the local interlayer synchronization error $E_k^{ij}$ for corresponding pairs of nodes $k$ from two layers:
\begin{equation}
    E_{k}^{ij}= \lim_{T \to \infty}\frac{1}{T}\int_0^T \!   \, \left\lVert \mathbf{x}_k^i(t)-\mathbf{x}_k^j(t)\right\rVert \mathrm{d}t. 
    \label{E_loc_eq}
\end{equation}
This measure uncovers complex regimes where only some of the oscillator pairs are synchronized, thus vanishing there.

Nonlocally coupled ring topologies within the layers allow for the observation of chimera states within the layers \cite{OME13}. An important feature of these patterns is the difference of mean phase velocities for synchronized and desynchronized groups of oscillators~\cite{KUR02a,ABR04}. Usually, oscillators in synchronized domains are phase-locked and have identical frequencies, while oscillators from the incoherent domain speed up (or slow down) and the mean frequency profile assumes an arc-like shape. The mean phase velocity for each oscillator is calculated as
\begin{align}
\omega_k = \frac{2\pi M_k}{\Delta T} \medspace \medspace , \medspace k=1,2,...N   ,
\label{omega_eq}
\end{align}
where $M_k$ is the number of complete oscillations of the $k$th node during the average time $\Delta T$.

\section{Results}
\begin{figure} %[h!]
\centering
\includegraphics[width=8.5cm]{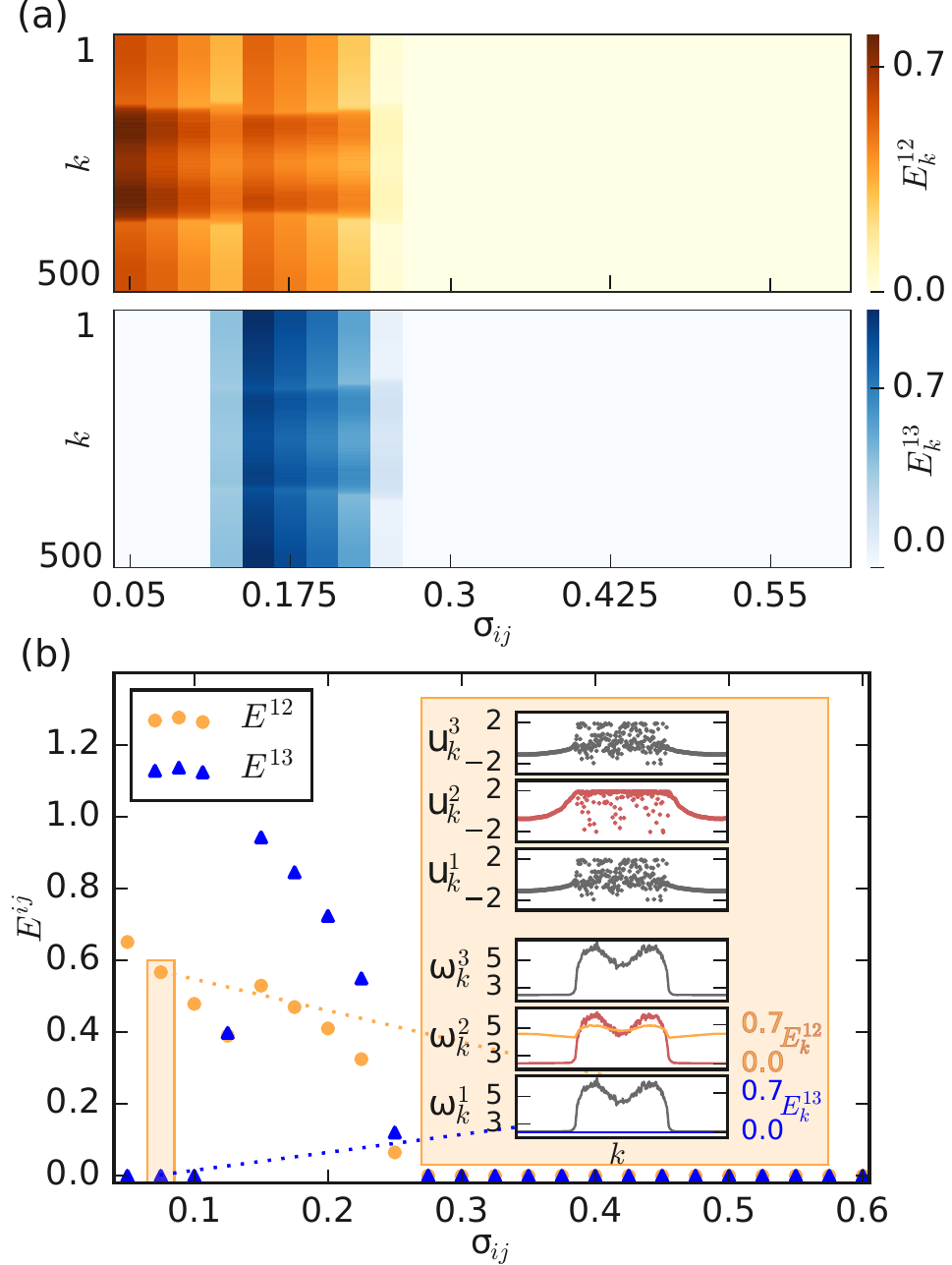}
\caption{(a) Local and (b) global synchronization error for averaged over 5 random initial conditions with identical layers versus interlayer coupling strength $\sigma_{ij}\equiv\sigma_{12}=\sigma_{32}$. One can distinguish three regions of dynamics with increasing $\sigma_{ij}$: (i) relay synchronization, (ii) desynchronization and (iii) full synchronization. The insets in (b) show snapshots $u_k^i$ and mean phase velocity profiles $\omega_k^i$ together with local synchronization errors $E_k^{12}$ (orange) and $E_k^{13}$ (blue) for a coupling strength of $\sigma_{ij}=0.075$. The relay layer is marked in red and the remote layers in grey. Parameters are $\epsilon =0.05$, $a=0.5$ $R=170$, $\phi=\frac{\pi}{2}-0.1$, $N=500$, $\sigma_i=0.2$ for layers $i=1,2,3$, $\sigma_{ij}\in [0.05,0.6]$. The simulation time is $t_{max}=2000$ with time steps of $\Delta t=0.05$.}

\label{Identical_layers}
\end{figure}

%\subsection{Identical Layers}
Before starting to introduce topological inhomogeneities in our network, we address the case of the unperturbed network where full and partial relay synchronization is observed \cite{SAW18c}. In contrast to \cite{SAW18c} the interlayer coupling delay is neglected. Figure~\ref{Identical_layers} displays averaged results of five numerical simulations for different sets of random initial conditions for the three-layer network Eq.~(\ref{dynamics_eq_end}). Figure~\ref{Identical_layers}(a) shows the local synchronization errors between the remote layers and the relay layer (upper panel, orange), and between the two remote layers (middle panel, blue), with increasing interlayer coupling strength. Figure~\ref{Identical_layers}(b) condenses those results by depicting the global synchronization errors. One can identify three characteristic regimes. In the first one, for small interlayer coupling strength, we observe relay synchronization with zero error $E^{13}$ between the remote layers $1$ and $3$, and nonzero error $E^{12}$ between the remote and relay layers. In the second regime, with further increase of $\sigma_{ij}\equiv\sigma_{12}=\sigma_{32}$, all three layers are desynchronized. Finally, for large interlayer coupling strength, in the third regime, the whole network is completely synchronized ($E^{12}=E^{13}=0$). The inset demonstrates an example of relay synchronization for $\sigma_{ij}=0.075$ (region shaded orange in panel (b)): upper panels depict snapshots of chimera states in three layers, and bottom panels show the corresponding mean phase velocity profiles $\omega_k^{i}$ together with local synchronization errors $E_k^{ij}$.

%\subsection{Non-Identical Layers}
In the following, we change the connectivities inside the network layers step by step towards a random topology following the Watts-Strogatz algorithm \cite{WAT98}. For this, we introduce the parameter $p$ which determines the probability for each link of the regular ring to be replaced by a random link, and therefore serves as a measure for the degree of randomness in the network. First, we investigate the influence of the parameter $p$ on the network properties.

\begin{figure} %[h!]
\centering
%\hspace{-1.5cm}
\includegraphics[width=9cm]{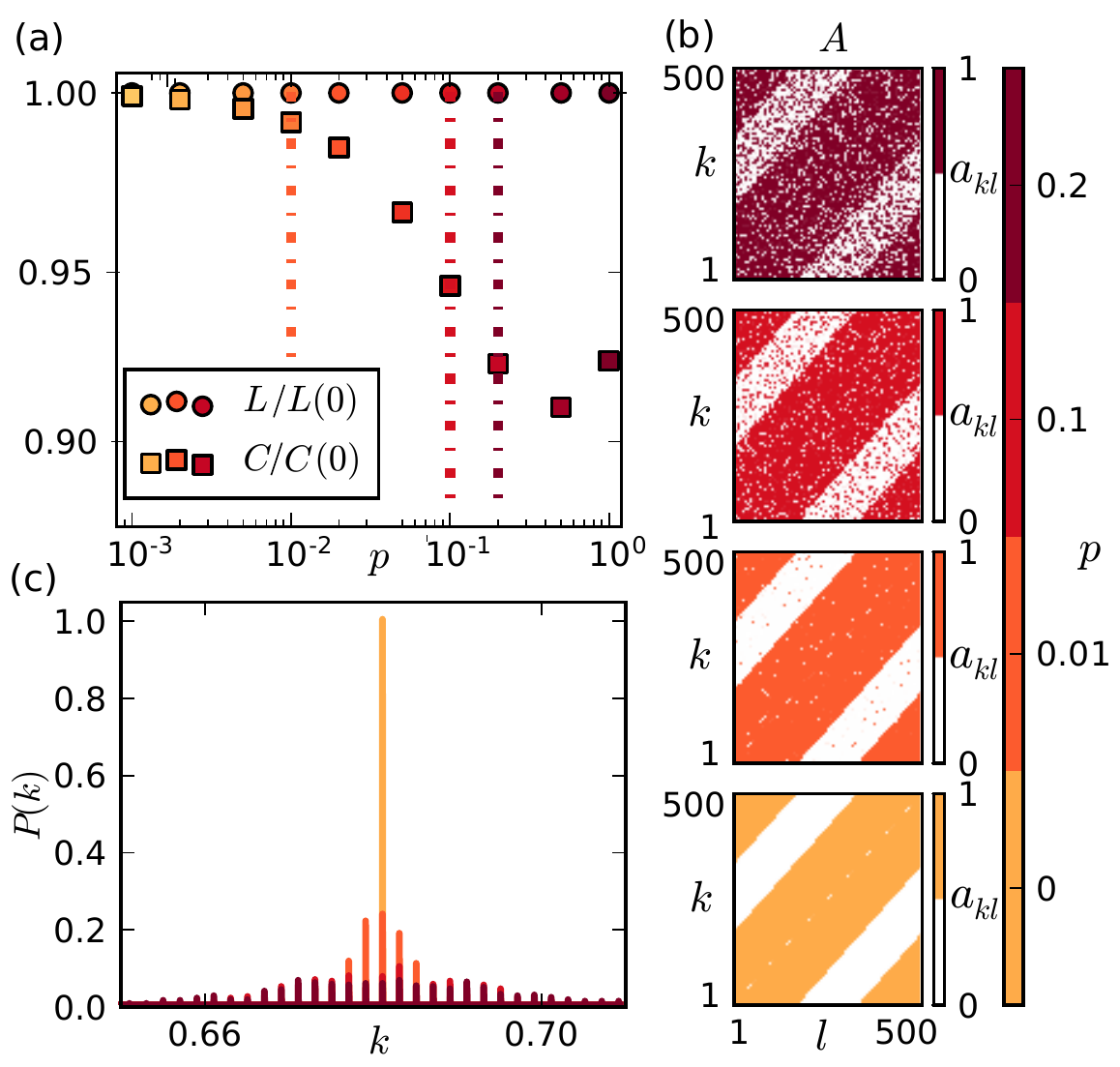}
\caption{
(a) Normalized shortest path length $L(p)/L(0)$ (circles) and normalized clustering coefficient $C(p)/C(0)$ (squares) averaged over 5 random realizations of the Watts-Strogatz connectivity for increasing rewiring probability $p$ (color coded, see color scale on the right), generated with the Watts-Strogatz algorithm~\cite{WAT98}.   
(b) $N \times N$-adjacency matrices $a_{kl}$ for different $p$ ($p=0,0.01,0.1,0.2$, see color scale on the right). 
(c) Histograms showing the normalized degree distribution $P(k)$ of a Watts-Strogatz graph with $N=500$, $R=170$ and increasing $p$ ($p=0,0.01,0.1,0.2$, see color scale on the right). 
Other parameters as in Fig.~\ref{Identical_layers}.}
\label{p_influence}
\end{figure}

With increasing probability of random shortcuts, we analyze the clustering coefficients $C$ and average shortest path length $L$ within the inhomogeneous layer. Figure~\ref{p_influence}(a) demonstrates that increasing the randomness $p$ results in a decreasing clustering coefficient, corresponding to the formation of low- and high-degree nodes (Fig.~\ref{p_influence}(c)) and thus to a growing topological inhomogeneity within one layer. Because of the small ensemble size used for averaging, artefacts may arise, e.g., the data point at $p=1$ does not continue to decrease. At the same time, we observe almost constant shortest path length in Fig.~\ref{p_influence}(a). Due to the large nonlocal coupling range $R$ the shortest path length is already quite small for the regular ring ($p=0$) and is not changed much by the introduction of shortcuts. This is in contrast to classical small-world networks which start rewiring from a sparse nonlocally coupled ring network (small coupling range), and hence exhibit a pronounced transition from large $C$ and $L$ (sparse ring) to small $C$ and $L$ (random network) via an intermediate small-world regime with large $C$ and small $L$. 
The adjacency matrices of a layer are shown in Fig.~\ref{p_influence}(b) for different $p$. For $p=0$ the matrix is symmetric and contains spatially separated homogeneous regions of coupling (colored) and no coupling (white), representing the regular nonlocally coupled ring. Increasing $p$ leads to the creation of short-cuts in the network which connect distant spatial domains. 
\\

\begin{figure} %[h!]
\centering
\includegraphics[width=8.5cm]{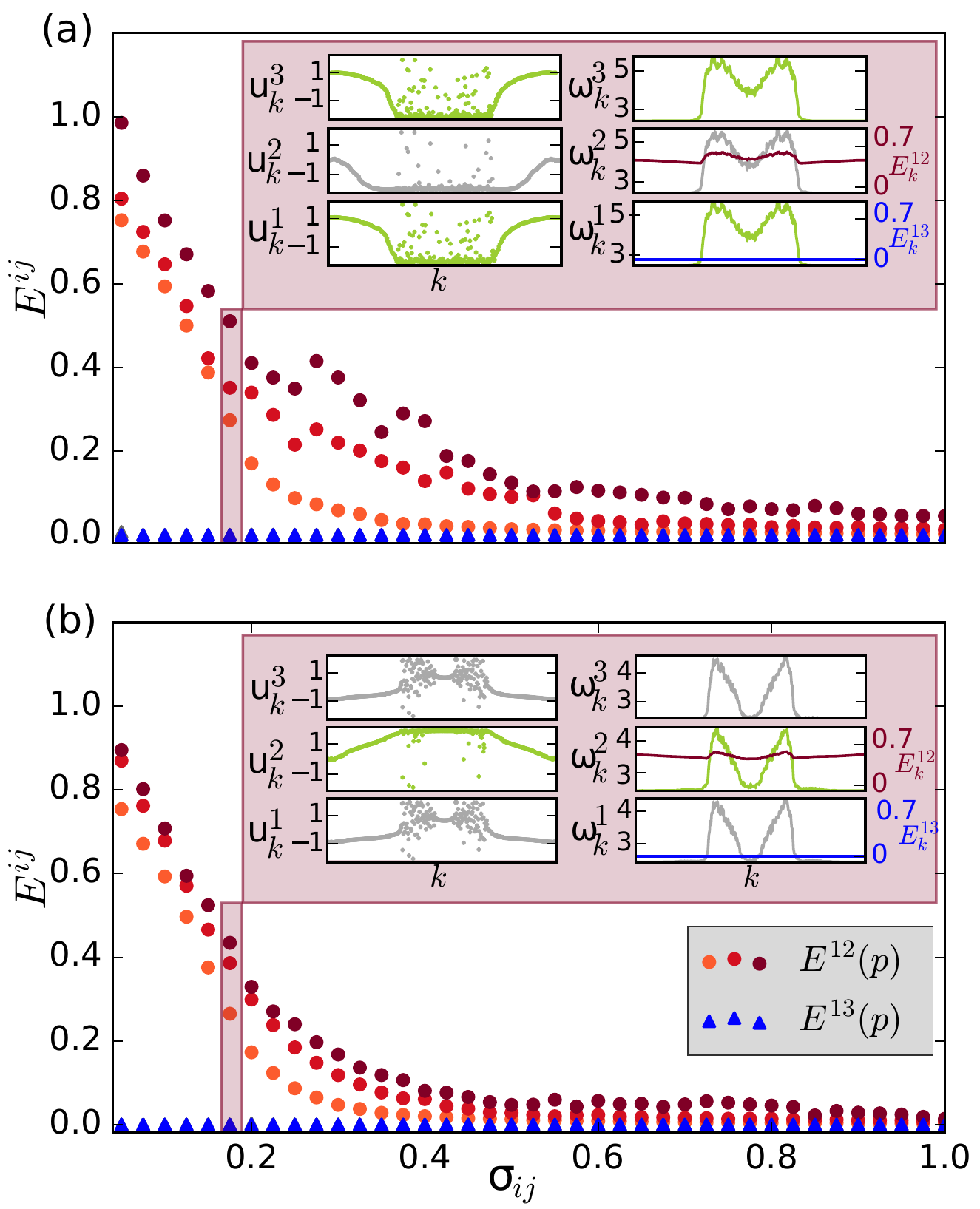}
\caption{Global synchronization error $E^{ij}$ versus coupling strength $\sigma_{ij}\equiv \sigma_{12}=\sigma_{32}$, where $E^{13}=0$ is shown by blue triangles, and $E^{12}$ is marked by colored circles for 3 different values of $p$ ($p = 0.01,0.1,0.2$, color-coded as in Fig.~\ref{p_influence}). For each value of $p$, we average over $5$ network realizations with different random sets of initial conditions. (a) Topology of the remote layers is modified, (b) topology of the relay layer is changed. The insets show snapshots $u_k^i$ and mean phase velocity profiles $\omega_k^i$ with local synchronization errors $E_k^{12}$ (red), $E_k^{13}$ (blue) for all three layers for $\sigma_{ij}=0.175$ and $p=0.2$, where the manipulated layers are marked green. Other parameters as in Fig.~\ref{Identical_layers}.}
\label{p_statistics}
\end{figure}
In order to examine the influence of $p$ upon the relay synchronization scenario, we focus on the %statistical representation of the
global synchronization error versus the interlayer coupling strength (Fig.~\ref{p_statistics}).
In the following, the topology of the layers is modified with the rewiring probability $p$ in two different ways: We either change the topology of the remote layers (applying the same random realization to both layers), or the relay layer, while keeping the regular nonlocally coupled ring structure in the relay layer or remote layers, respectively.

Our simulations with random initial conditions show that besides the familiar scenario (relay interlayer synchronization - desynchronization - full interlayer synchronization) observed for the unperturbed three-layer network, frequently a different scenario occurs, where the system changes its dynamics directly from relay to full synchronization without being desynchronized in between. This new transition is shown in Fig.~\ref{p_statistics}.
For both cases of inhomogeneity increasing $p$ leads to a growing global synchronization error $E^{12}$, such that the threshold value of interlayer coupling $\sigma_{ij}^{*}$, up to which the system remains relay-synchronized, before becoming fully synchronized, shifts substantially towards higher values. Snapshots and mean phase velocity profiles with local synchronization errors are shown in the inset, depicting the relay synchronization of chimera states for an interlayer coupling of $\sigma_{ij}=0.175$ and $p=0.2$. Here, we further observe the emergence of a two-headed chimera state, i.e., two incoherent domains, in the case of an inhomogeneous relay layer. In case of inhomogeneous remote layers the two-headed chimera state is not observed. As it is known from Ref.~\onlinecite{OME13}, two-headed chimera states emerge due to a subtle interplay between coupling strength and nonlocality in the coupling structure. Remarkably, the relay layer is more sensitive to the introduction of nonlocality in the sense of Watts and Strogatz. 

\begin{figure} %[h!]
	\centering
	%\begin{minipage}{1\textwidth}
	\includegraphics[width=8cm]{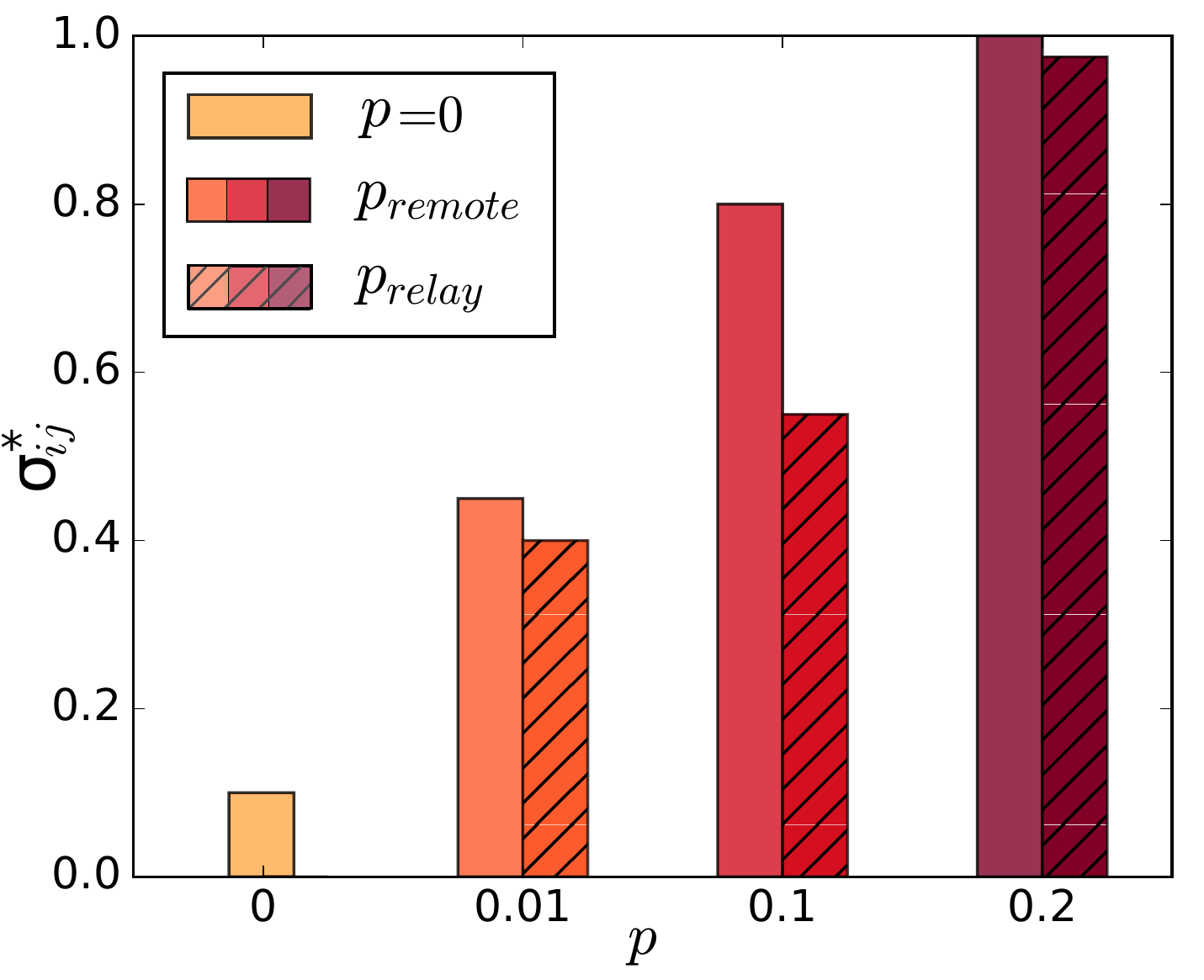}
	\caption{Threshold value $\sigma_{ij}^*$ up to which the system remains relay-synchronized depending on the interlayer coupling strength $\sigma_{ij}$ and the link rewiring probability $p=0,0.01,0.1,0.2$. The values $\sigma_{ij}^*(p)$ are calculated from the data of Fig.~\ref{p_statistics} where the system is assumed to be fully synchronized if $E^{12} \leq 0.02$. The left and right bars refer to inhomogeneous remote layers (plain) or relay layer (hatched), respectively.}
	\label{ges_stat}
	%\end{minipage}
\end{figure} 
Figure~\ref{ges_stat} illustrates the results of our numerical simulations for three non-zero values of the link rewiring probability $p$. It compares the critical interlayer coupling strength $\sigma_{ij}^*$ at the transition from relay synchronization to complete synchronization, and for comparison also gives the threshold of relay synchronization for $p=0$. We find that networks with inhomogeneous remote layers have higher thresholds, while the transition to complete synchronization for a network with an inhomogeneous relay layer takes place for slightly lower coupling strengths. Our analysis shows that introducing randomness in the topology of either the remote or the relay layer has an advantageous effect upon the robustness of the regime of relay synchronization in the three-layer network. We observe that with increasing rewiring probability the threshold values of the interlayer coupling strength, up to which the relay-synchronized state is preserved, increase significantly.

\section{Conclusion}

In summary, we have analyzed nontrivial synchronization scenarios in three-layer networks of FitzHugh-Nagumo oscillators with inhomogeneous topologies constructed by a small-world algorithm. In a triplex network with identical homogeneous layers, varying the interlayer coupling strength allows for regimes of relay synchronization, desynchronization, and complete synchronization. To check the robustness of these regimes, we have introduced topological inhomogeneities in the remote layers or the relay layer by randomly rewiring existing regular links and replacing them with random shortcuts with a probability $p$. 

Complementary to the known scenarios for a regular nonlocally coupled ring network, we find a novel scenario with a direct transition from the regime of relay synchronization to the regime of complete synchronization without a desynchronized regime in between. This observation suggests that the introduction of inhomogeneities generally supports synchronization. Moreover, these findings are in agreement with results for one-layer networks~\cite{ARE08} showing that regular ring networks exhibit poor synchronizability, which is significantly improved when random connections are added~\cite{BAR02} or links are rewired within the network~\cite{HON04}. Thus we have extended those results from one-layer towards multilayer systems and even generalized them for partial synchronization patterns such as chimera states.

Our findings demonstrate that random inhomogeneities in the network topology have a positive effect on the relay synchronization by increasing the parameter range where it is observed. This effect increases strongly with increasing link rewiring probability $p$. Comparing two cases of topological inhomogeneities, i.e., modification of the remote layers or the relay layer, respectively, we conclude that inhomogeneity in the remote layers is more advantageous for relay synchronization. In this case, relay synchronization can be observed for a wider range of the interlayer coupling strength. 

Keeping the coupling range $R$ in the initial network fixed and relatively large within the layers allows for the observation of chimera states with one incoherent domain in each layer. The case of more diluted topologies, i.e., a nonlocally coupled ring with a smaller coupling range, would result in chimera-like patterns with a larger number of alternating coherent and incoherent domains. But even for the large $R$ we have used, an inhomogeneous relay layer can induce chimeras with two incoherent domains in all three layers.\\

\begin{acknowledgments}
This work was supported by the Deutsche Forschungsgemeinschaft (DFG, German Research Foundation) - Project Nos. - 163436311 - SFB 910, 411803875 and 308748074.%Projects SCHO 307/15-1 and YA 225/3-1.
\end{acknowledgments}

\section*{Data Availability Statement}
The data that supports the findings of this study are available within the article.


\begin{thebibliography}{59}%
	\makeatletter
	\providecommand \@ifxundefined [1]{%
		\@ifx{#1\undefined}
	}%
	\providecommand \@ifnum [1]{%
		\ifnum #1\expandafter \@firstoftwo
		\else \expandafter \@secondoftwo
		\fi
	}%
	\providecommand \@ifx [1]{%
		\ifx #1\expandafter \@firstoftwo
		\else \expandafter \@secondoftwo
		\fi
	}%
	\providecommand \natexlab [1]{#1}%
	\providecommand \enquote  [1]{``#1''}%
	\providecommand \bibnamefont  [1]{#1}%
	\providecommand \bibfnamefont [1]{#1}%
	\providecommand \citenamefont [1]{#1}%
	\providecommand \href@noop [0]{\@secondoftwo}%
	\providecommand \href [0]{\begingroup \@sanitize@url \@href}%
	\providecommand \@href[1]{\@@startlink{#1}\@@href}%
	\providecommand \@@href[1]{\endgroup#1\@@endlink}%
	\providecommand \@sanitize@url [0]{\catcode `\\12\catcode `\$12\catcode
		`\&12\catcode `\#12\catcode `\^12\catcode `\_12\catcode `\%12\relax}%
	\providecommand \@@startlink[1]{}%
	\providecommand \@@endlink[0]{}%
	\providecommand \url  [0]{\begingroup\@sanitize@url \@url }%
	\providecommand \@url [1]{\endgroup\@href {#1}{\urlprefix }}%
	\providecommand \urlprefix  [0]{URL }%
	\providecommand \Eprint [0]{\href }%
	\providecommand \doibase [0]{http://dx.doi.org/}%
	\providecommand \selectlanguage [0]{\@gobble}%
	\providecommand \bibinfo  [0]{\@secondoftwo}%
	\providecommand \bibfield  [0]{\@secondoftwo}%
	\providecommand \translation [1]{[#1]}%
	\providecommand \BibitemOpen [0]{}%
	\providecommand \bibitemStop [0]{}%
	\providecommand \bibitemNoStop [0]{.\EOS\space}%
	\providecommand \EOS [0]{\spacefactor3000\relax}%
	\providecommand \BibitemShut  [1]{\csname bibitem#1\endcsname}%
	\let\auto@bib@innerbib\@empty
	%</preamble>
	\bibitem [{\citenamefont {Pikovsky}\ \emph {et~al.}(2001)\citenamefont
		{Pikovsky}, \citenamefont {Rosenblum},\ and\ \citenamefont {Kurths}}]{PIK01}%
	\BibitemOpen
	\bibfield  {author} {\bibinfo {author} {\bibfnamefont {A.}~\bibnamefont
			{Pikovsky}}, \bibinfo {author} {\bibfnamefont {M.}~\bibnamefont {Rosenblum}},
		\ and\ \bibinfo {author} {\bibfnamefont {J.}~\bibnamefont {Kurths}},\
	}\href@noop {} {\emph {\bibinfo {title} {Synchronization: a universal concept
				in nonlinear sciences}}}\ (\bibinfo  {publisher} {Cambridge University
		Press},\ \bibinfo {address} {Cambridge},\ \bibinfo {year} {2001})\BibitemShut
	{NoStop}%
	\bibitem [{\citenamefont {Boccaletti}\ \emph {et~al.}(2014)\citenamefont
		{Boccaletti}, \citenamefont {Bianconi}, \citenamefont {Criado}, \citenamefont
		{del Genio}, \citenamefont {G\'omez-Garde\~nes}, \citenamefont {Romance},
		\citenamefont {Sendi\~{n}a Nadal}, \citenamefont {Wang},\ and\ \citenamefont
		{Zanin}}]{BOC14}%
	\BibitemOpen
	\bibfield  {author} {\bibinfo {author} {\bibfnamefont {S.}~\bibnamefont
			{Boccaletti}}, \bibinfo {author} {\bibfnamefont {G.}~\bibnamefont
			{Bianconi}}, \bibinfo {author} {\bibfnamefont {R.}~\bibnamefont {Criado}},
		\bibinfo {author} {\bibfnamefont {C.~I.}\ \bibnamefont {del Genio}}, \bibinfo
		{author} {\bibfnamefont {J.}~\bibnamefont {G\'omez-Garde\~nes}}, \bibinfo
		{author} {\bibfnamefont {M.}~\bibnamefont {Romance}}, \bibinfo {author}
		{\bibfnamefont {I.}~\bibnamefont {Sendi\~{n}a Nadal}}, \bibinfo {author}
		{\bibfnamefont {Z.}~\bibnamefont {Wang}}, \ and\ \bibinfo {author}
		{\bibfnamefont {M.}~\bibnamefont {Zanin}},\ }\href {\doibase
		10.1016/j.physrep.2014.07.001} {\bibfield  {journal} {\bibinfo  {journal}
			{Phys. Rep.}\ }\textbf {\bibinfo {volume} {544}},\ \bibinfo {pages} {1}
		(\bibinfo {year} {2014})}\BibitemShut {NoStop}%
	\bibitem [{\citenamefont {Gollo}\ \emph {et~al.}(2011)\citenamefont {Gollo},
		\citenamefont {Mirasso}, \citenamefont {Atienza}, \citenamefont
		{Crespo-Garcia},\ and\ \citenamefont {Cantero}}]{GOL11a}%
	\BibitemOpen
	\bibfield  {author} {\bibinfo {author} {\bibfnamefont {L.~L.}\ \bibnamefont
			{Gollo}}, \bibinfo {author} {\bibfnamefont {C.~R.}\ \bibnamefont {Mirasso}},
		\bibinfo {author} {\bibfnamefont {M.}~\bibnamefont {Atienza}}, \bibinfo
		{author} {\bibfnamefont {M.}~\bibnamefont {Crespo-Garcia}}, \ and\ \bibinfo
		{author} {\bibfnamefont {J.~L.}\ \bibnamefont {Cantero}},\ }\href@noop {}
	{\bibfield  {journal} {\bibinfo  {journal} {PLoS ONE}\ }\textbf {\bibinfo
			{volume} {6}},\ \bibinfo {pages} {e17756} (\bibinfo {year}
		{2011})}\BibitemShut {NoStop}%
	\bibitem [{\citenamefont {Fischer}\ \emph {et~al.}(2006)\citenamefont
		{Fischer}, \citenamefont {Vicente}, \citenamefont {Buld{\'u}}, \citenamefont
		{Peil}, \citenamefont {Mirasso}, \citenamefont {Torrent},\ and\ \citenamefont
		{Garc{\'i}a-Ojalvo}}]{FIS06}%
	\BibitemOpen
	\bibfield  {author} {\bibinfo {author} {\bibfnamefont {I.}~\bibnamefont
			{Fischer}}, \bibinfo {author} {\bibfnamefont {R.}~\bibnamefont {Vicente}},
		\bibinfo {author} {\bibfnamefont {J.~M.}\ \bibnamefont {Buld{\'u}}}, \bibinfo
		{author} {\bibfnamefont {M.}~\bibnamefont {Peil}}, \bibinfo {author}
		{\bibfnamefont {C.~R.}\ \bibnamefont {Mirasso}}, \bibinfo {author}
		{\bibfnamefont {M.~C.}\ \bibnamefont {Torrent}}, \ and\ \bibinfo {author}
		{\bibfnamefont {J.}~\bibnamefont {Garc{\'i}a-Ojalvo}},\ }\href {\doibase
		10.1103/physrevlett.97.123902} {\bibfield  {journal} {\bibinfo  {journal}
			{Phys. Rev. Lett.}\ }\textbf {\bibinfo {volume} {97}},\ \bibinfo {pages}
		{123902} (\bibinfo {year} {2006})}\BibitemShut {NoStop}%
	\bibitem [{\citenamefont {Bergner}\ \emph {et~al.}(2012)\citenamefont
		{Bergner}, \citenamefont {Frasca}, \citenamefont {Sciuto}, \citenamefont
		{Buscarino}, \citenamefont {Ngamga}, \citenamefont {Fortuna},\ and\
		\citenamefont {Kurths}}]{BER12}%
	\BibitemOpen
	\bibfield  {author} {\bibinfo {author} {\bibfnamefont {A.}~\bibnamefont
			{Bergner}}, \bibinfo {author} {\bibfnamefont {M.}~\bibnamefont {Frasca}},
		\bibinfo {author} {\bibfnamefont {G.}~\bibnamefont {Sciuto}}, \bibinfo
		{author} {\bibfnamefont {A.}~\bibnamefont {Buscarino}}, \bibinfo {author}
		{\bibfnamefont {E.~J.}\ \bibnamefont {Ngamga}}, \bibinfo {author}
		{\bibfnamefont {L.}~\bibnamefont {Fortuna}}, \ and\ \bibinfo {author}
		{\bibfnamefont {J.}~\bibnamefont {Kurths}},\ }\href {\doibase
		10.1103/physreve.85.026208} {\bibfield  {journal} {\bibinfo  {journal} {Phys.
				Rev. E}\ }\textbf {\bibinfo {volume} {85}},\ \bibinfo {pages} {026208}
		(\bibinfo {year} {2012})}\BibitemShut {NoStop}%
	\bibitem [{\citenamefont {Leyva}\ \emph {et~al.}(2018)\citenamefont {Leyva},
		\citenamefont {Sendi{\~n}a-Nadal}, \citenamefont {Sevilla-Escoboza},
		\citenamefont {Vera-Avila}, \citenamefont {Chholak},\ and\ \citenamefont
		{Boccaletti}}]{LEY18}%
	\BibitemOpen
	\bibfield  {author} {\bibinfo {author} {\bibfnamefont {I.}~\bibnamefont
			{Leyva}}, \bibinfo {author} {\bibfnamefont {I.}~\bibnamefont
			{Sendi{\~n}a-Nadal}}, \bibinfo {author} {\bibfnamefont {R.}~\bibnamefont
			{Sevilla-Escoboza}}, \bibinfo {author} {\bibfnamefont {V.~P.}\ \bibnamefont
			{Vera-Avila}}, \bibinfo {author} {\bibfnamefont {P.}~\bibnamefont {Chholak}},
		\ and\ \bibinfo {author} {\bibfnamefont {S.}~\bibnamefont {Boccaletti}},\
	}\href@noop {} {\bibfield  {journal} {\bibinfo  {journal} {Sci. Rep.}\
		}\textbf {\bibinfo {volume} {8}},\ \bibinfo {pages} {8629} (\bibinfo {year}
		{2018})}\BibitemShut {NoStop}%
	\bibitem [{\citenamefont {Sawicki}\ \emph {et~al.}(2018)\citenamefont
		{Sawicki}, \citenamefont {Omelchenko}, \citenamefont {Zakharova},\ and\
		\citenamefont {Sch{\"o}ll}}]{SAW18c}%
	\BibitemOpen
	\bibfield  {author} {\bibinfo {author} {\bibfnamefont {J.}~\bibnamefont
			{Sawicki}}, \bibinfo {author} {\bibfnamefont {I.}~\bibnamefont {Omelchenko}},
		\bibinfo {author} {\bibfnamefont {A.}~\bibnamefont {Zakharova}}, \ and\
		\bibinfo {author} {\bibfnamefont {E.}~\bibnamefont {Sch{\"o}ll}},\
	}\href@noop {} {\bibfield  {journal} {\bibinfo  {journal} {Phys. Rev. E}\
		}\textbf {\bibinfo {volume} {98}},\ \bibinfo {pages} {062224} (\bibinfo
		{year} {2018})}\BibitemShut {NoStop}%
	\bibitem [{\citenamefont {Girvan}\ and\ \citenamefont {Newman}(2002)}]{GIR02}%
	\BibitemOpen
	\bibfield  {author} {\bibinfo {author} {\bibfnamefont {M.}~\bibnamefont
			{Girvan}}\ and\ \bibinfo {author} {\bibfnamefont {M.~E.~J.}\ \bibnamefont
			{Newman}},\ }\href@noop {} {\bibfield  {journal} {\bibinfo  {journal} {Proc.
				Natl. Acad. Sci. USA}\ }\textbf {\bibinfo {volume} {99}},\ \bibinfo {pages}
		{7821} (\bibinfo {year} {2002})}\BibitemShut {NoStop}%
	\bibitem [{\citenamefont {Moslonka-Lefebvre}\ \emph {et~al.}(2016)\citenamefont
		{Moslonka-Lefebvre}, \citenamefont {Gilligan}, \citenamefont {Monod},
		\citenamefont {Belloc}, \citenamefont {Ezanno}, \citenamefont {Filipe},\ and\
		\citenamefont {Vergu}}]{MOS16}%
	\BibitemOpen
	\bibfield  {author} {\bibinfo {author} {\bibfnamefont {M.}~\bibnamefont
			{Moslonka-Lefebvre}}, \bibinfo {author} {\bibfnamefont {C.~A.}\ \bibnamefont
			{Gilligan}}, \bibinfo {author} {\bibfnamefont {H.}~\bibnamefont {Monod}},
		\bibinfo {author} {\bibfnamefont {C.}~\bibnamefont {Belloc}}, \bibinfo
		{author} {\bibfnamefont {P.}~\bibnamefont {Ezanno}}, \bibinfo {author}
		{\bibfnamefont {J.~A.~N.}\ \bibnamefont {Filipe}}, \ and\ \bibinfo {author}
		{\bibfnamefont {E.}~\bibnamefont {Vergu}},\ }\href@noop {} {\bibfield
		{journal} {\bibinfo  {journal} {Journal of The Royal Society Interface}\
		}\textbf {\bibinfo {volume} {13}},\ \bibinfo {pages} {20151099} (\bibinfo
		{year} {2016})}\BibitemShut {NoStop}%
	\bibitem [{\citenamefont {Grimm}\ \emph {et~al.}(2005)\citenamefont {Grimm},
		\citenamefont {Revilla}, \citenamefont {Berger} \emph {et~al.}}]{GRI05}%
	\BibitemOpen
	\bibfield  {author} {\bibinfo {author} {\bibfnamefont {V.}~\bibnamefont
			{Grimm}}, \bibinfo {author} {\bibfnamefont {E.}~\bibnamefont {Revilla}},
		\bibinfo {author} {\bibfnamefont {U.}~\bibnamefont {Berger}},  \emph
		{et~al.},\ }\href {\doibase 10.1126/science.1116681} {\bibfield  {journal}
		{\bibinfo  {journal} {Science}\ }\textbf {\bibinfo {volume} {310}},\ \bibinfo
		{pages} {987} (\bibinfo {year} {2005})}\BibitemShut {NoStop}%
	\bibitem [{\citenamefont {Sprott}(2004)}]{SPR04a}%
	\BibitemOpen
	\bibfield  {author} {\bibinfo {author} {\bibfnamefont {J.~C.}\ \bibnamefont
			{Sprott}},\ }\href@noop {} {\bibfield  {journal} {\bibinfo  {journal} {Phys.
				Lett. A}\ }\textbf {\bibinfo {volume} {325}},\ \bibinfo {pages} {329}
		(\bibinfo {year} {2004})}\BibitemShut {NoStop}%
	\bibitem [{\citenamefont {Woolley-Meza}\ \emph {et~al.}(2011)\citenamefont
		{Woolley-Meza}, \citenamefont {Thiemann}, \citenamefont {Grady},
		\citenamefont {Lee}, \citenamefont {Seebens}, \citenamefont {Blasius},\ and\
		\citenamefont {Brockmann}}]{WOO11}%
	\BibitemOpen
	\bibfield  {author} {\bibinfo {author} {\bibfnamefont {O.}~\bibnamefont
			{Woolley-Meza}}, \bibinfo {author} {\bibfnamefont {C.}~\bibnamefont
			{Thiemann}}, \bibinfo {author} {\bibfnamefont {D.}~\bibnamefont {Grady}},
		\bibinfo {author} {\bibfnamefont {J.~J.}\ \bibnamefont {Lee}}, \bibinfo
		{author} {\bibfnamefont {H.}~\bibnamefont {Seebens}}, \bibinfo {author}
		{\bibfnamefont {B.}~\bibnamefont {Blasius}}, \ and\ \bibinfo {author}
		{\bibfnamefont {D.}~\bibnamefont {Brockmann}},\ }\href {\doibase
		10.1140/epjb/e2011-20208-9} {\bibfield  {journal} {\bibinfo  {journal} {Eur.
				Phys. J. B}\ }\textbf {\bibinfo {volume} {84}},\ \bibinfo {pages} {589}
		(\bibinfo {year} {2011})}\BibitemShut {NoStop}%
	\bibitem [{\citenamefont {Cardillo}\ \emph {et~al.}(2013)\citenamefont
		{Cardillo}, \citenamefont {Zanin}, \citenamefont {G{\`o}mez Garde\~nes},
		\citenamefont {Romance}, \citenamefont {Garcia~del Amo},\ and\ \citenamefont
		{Boccaletti}}]{CAR13d}%
	\BibitemOpen
	\bibfield  {author} {\bibinfo {author} {\bibfnamefont {A.}~\bibnamefont
			{Cardillo}}, \bibinfo {author} {\bibfnamefont {M.}~\bibnamefont {Zanin}},
		\bibinfo {author} {\bibfnamefont {J.}~\bibnamefont {G{\`o}mez Garde\~nes}},
		\bibinfo {author} {\bibfnamefont {M.}~\bibnamefont {Romance}}, \bibinfo
		{author} {\bibfnamefont {A.}~\bibnamefont {Garcia~del Amo}}, \ and\ \bibinfo
		{author} {\bibfnamefont {S.}~\bibnamefont {Boccaletti}},\ }\href@noop {}
	{\bibfield  {journal} {\bibinfo  {journal} {Eur. Phys. J. ST}\ }\textbf
		{\bibinfo {volume} {215}},\ \bibinfo {pages} {23} (\bibinfo {year}
		{2013})}\BibitemShut {NoStop}%
	\bibitem [{\citenamefont {Bentley}\ \emph {et~al.}(2016)\citenamefont
		{Bentley}, \citenamefont {Branicky}, \citenamefont {Barnes}, \citenamefont
		{Chew}, \citenamefont {Yemini}, \citenamefont {Bullmore}, \citenamefont
		{V{\'e}tes},\ and\ \citenamefont {Schafer}}]{BEN16}%
	\BibitemOpen
	\bibfield  {author} {\bibinfo {author} {\bibfnamefont {B.}~\bibnamefont
			{Bentley}}, \bibinfo {author} {\bibfnamefont {R.}~\bibnamefont {Branicky}},
		\bibinfo {author} {\bibfnamefont {C.~L.}\ \bibnamefont {Barnes}}, \bibinfo
		{author} {\bibfnamefont {Y.~L.}\ \bibnamefont {Chew}}, \bibinfo {author}
		{\bibfnamefont {E.}~\bibnamefont {Yemini}}, \bibinfo {author} {\bibfnamefont
			{E.~T.}\ \bibnamefont {Bullmore}}, \bibinfo {author} {\bibfnamefont {P.~E.}\
			\bibnamefont {V{\'e}tes}}, \ and\ \bibinfo {author} {\bibfnamefont {W.~R.}\
			\bibnamefont {Schafer}},\ }\href {\doibase 10.1371/journal.pcbi.1005283}
	{\bibfield  {journal} {\bibinfo  {journal} {PLOS Comput. Biol.}\ }\textbf
		{\bibinfo {volume} {12}},\ \bibinfo {pages} {e1005283} (\bibinfo {year}
		{2016})}\BibitemShut {NoStop}%
	\bibitem [{\citenamefont {Battiston}\ \emph {et~al.}(2017)\citenamefont
		{Battiston}, \citenamefont {Nicosia}, \citenamefont {Chavez},\ and\
		\citenamefont {Latora}}]{BAT17}%
	\BibitemOpen
	\bibfield  {author} {\bibinfo {author} {\bibfnamefont {F.}~\bibnamefont
			{Battiston}}, \bibinfo {author} {\bibfnamefont {V.}~\bibnamefont {Nicosia}},
		\bibinfo {author} {\bibfnamefont {M.}~\bibnamefont {Chavez}}, \ and\ \bibinfo
		{author} {\bibfnamefont {V.}~\bibnamefont {Latora}},\ }\href@noop {}
	{\bibfield  {journal} {\bibinfo  {journal} {Chaos}\ }\textbf {\bibinfo
			{volume} {27}},\ \bibinfo {pages} {047404} (\bibinfo {year}
		{2017})}\BibitemShut {NoStop}%
	\bibitem [{\citenamefont {Chouzouris}\ \emph {et~al.}(2018)\citenamefont
		{Chouzouris}, \citenamefont {Omelchenko}, \citenamefont {Zakharova},
		\citenamefont {Hlinka}, \citenamefont {Jiruska},\ and\ \citenamefont
		{Sch{\"o}ll}}]{CHO18}%
	\BibitemOpen
	\bibfield  {author} {\bibinfo {author} {\bibfnamefont {T.}~\bibnamefont
			{Chouzouris}}, \bibinfo {author} {\bibfnamefont {I.}~\bibnamefont
			{Omelchenko}}, \bibinfo {author} {\bibfnamefont {A.}~\bibnamefont
			{Zakharova}}, \bibinfo {author} {\bibfnamefont {J.}~\bibnamefont {Hlinka}},
		\bibinfo {author} {\bibfnamefont {P.}~\bibnamefont {Jiruska}}, \ and\
		\bibinfo {author} {\bibfnamefont {E.}~\bibnamefont {Sch{\"o}ll}},\ }\href
	{\doibase https://doi.org/10.1063/1.5009812} {\bibfield  {journal} {\bibinfo
			{journal} {Chaos}\ }\textbf {\bibinfo {volume} {28}},\ \bibinfo {pages}
		{045112} (\bibinfo {year} {2018})}\BibitemShut {NoStop}%
	\bibitem [{\citenamefont {Ramlow}\ \emph {et~al.}(2019)\citenamefont {Ramlow},
		\citenamefont {Sawicki}, \citenamefont {Zakharova}, \citenamefont {Hlinka},
		\citenamefont {Claussen},\ and\ \citenamefont {Sch{\"o}ll}}]{RAM19}%
	\BibitemOpen
	\bibfield  {author} {\bibinfo {author} {\bibfnamefont {L.}~\bibnamefont
			{Ramlow}}, \bibinfo {author} {\bibfnamefont {J.}~\bibnamefont {Sawicki}},
		\bibinfo {author} {\bibfnamefont {A.}~\bibnamefont {Zakharova}}, \bibinfo
		{author} {\bibfnamefont {J.}~\bibnamefont {Hlinka}}, \bibinfo {author}
		{\bibfnamefont {J.~C.}\ \bibnamefont {Claussen}}, \ and\ \bibinfo {author}
		{\bibfnamefont {E.}~\bibnamefont {Sch{\"o}ll}},\ }\href@noop {} {\bibfield
		{journal} {\bibinfo  {journal} {EPL}\ }\textbf {\bibinfo {volume} {126}},\
		\bibinfo {pages} {50007} (\bibinfo {year} {2019})},\ \bibinfo {note}
	{highlighted in phys.org
		https://phys.org/news/2019-07-unihemispheric-humans.html and in Europhys.
		News 50, no. 5-6 (2019)}\BibitemShut {NoStop}%
	\bibitem [{\citenamefont {Kivel{\"a}}\ \emph {et~al.}(2014)\citenamefont
		{Kivel{\"a}}, \citenamefont {Arenas}, \citenamefont {Barth{\'e}lemy},
		\citenamefont {Gleeson}, \citenamefont {Moreno},\ and\ \citenamefont
		{Porter}}]{KIV14}%
	\BibitemOpen
	\bibfield  {author} {\bibinfo {author} {\bibfnamefont {M.}~\bibnamefont
			{Kivel{\"a}}}, \bibinfo {author} {\bibfnamefont {A.}~\bibnamefont {Arenas}},
		\bibinfo {author} {\bibfnamefont {M.}~\bibnamefont {Barth{\'e}lemy}},
		\bibinfo {author} {\bibfnamefont {J.~P.}\ \bibnamefont {Gleeson}}, \bibinfo
		{author} {\bibfnamefont {Y.}~\bibnamefont {Moreno}}, \ and\ \bibinfo {author}
		{\bibfnamefont {M.~A.}\ \bibnamefont {Porter}},\ }\href {\doibase
		10.1093/comnet/cnu016} {\bibfield  {journal} {\bibinfo  {journal} {J. Complex
				Networks}\ }\textbf {\bibinfo {volume} {2}},\ \bibinfo {pages} {203}
		(\bibinfo {year} {2014})},\ \Eprint
	{http://arxiv.org/abs/http://comnet.oxfordjournals.org/content/2/3/203.full.pdf+html}
	{http://comnet.oxfordjournals.org/content/2/3/203.full.pdf+html} \BibitemShut
	{NoStop}%
	\bibitem [{\citenamefont {De~Domenico}\ \emph {et~al.}(2016)\citenamefont
		{De~Domenico}, \citenamefont {Granell}, \citenamefont {Porter},\ and\
		\citenamefont {Arenas}}]{DOM16a}%
	\BibitemOpen
	\bibfield  {author} {\bibinfo {author} {\bibfnamefont {M.}~\bibnamefont
			{De~Domenico}}, \bibinfo {author} {\bibfnamefont {C.}~\bibnamefont
			{Granell}}, \bibinfo {author} {\bibfnamefont {M.~A.}\ \bibnamefont {Porter}},
		\ and\ \bibinfo {author} {\bibfnamefont {A.}~\bibnamefont {Arenas}},\ }\href
	{\doibase 10.1038/nphys3865} {\bibfield  {journal} {\bibinfo  {journal} {Nat.
				Phys.}\ }\textbf {\bibinfo {volume} {12}},\ \bibinfo {pages} {901} (\bibinfo
		{year} {2016})}\BibitemShut {NoStop}%
	\bibitem [{\citenamefont {Boccaletti}\ \emph {et~al.}(2018)\citenamefont
		{Boccaletti}, \citenamefont {Pisarchik}, \citenamefont {del Genio},\ and\
		\citenamefont {Amann}}]{BOC18}%
	\BibitemOpen
	\bibfield  {author} {\bibinfo {author} {\bibfnamefont {S.}~\bibnamefont
			{Boccaletti}}, \bibinfo {author} {\bibfnamefont {A.~N.}\ \bibnamefont
			{Pisarchik}}, \bibinfo {author} {\bibfnamefont {C.~I.}\ \bibnamefont {del
				Genio}}, \ and\ \bibinfo {author} {\bibfnamefont {A.}~\bibnamefont {Amann}},\
	}\href@noop {} {\emph {\bibinfo {title} {Synchronization: From Coupled
				Systems to Complex Networks}}}\ (\bibinfo  {publisher} {Cambridge University
		Press},\ \bibinfo {address} {Cambridge},\ \bibinfo {year} {2018})\BibitemShut
	{NoStop}%
	\bibitem [{\citenamefont {Mucha}\ \emph {et~al.}(2010)\citenamefont {Mucha},
		\citenamefont {Richardson}, \citenamefont {Macon}, \citenamefont {Porter},\
		and\ \citenamefont {Onnela}}]{MUC10}%
	\BibitemOpen
	\bibfield  {author} {\bibinfo {author} {\bibfnamefont {P.~J.}\ \bibnamefont
			{Mucha}}, \bibinfo {author} {\bibfnamefont {T.}~\bibnamefont {Richardson}},
		\bibinfo {author} {\bibfnamefont {K.}~\bibnamefont {Macon}}, \bibinfo
		{author} {\bibfnamefont {M.~A.}\ \bibnamefont {Porter}}, \ and\ \bibinfo
		{author} {\bibfnamefont {J.~P.}\ \bibnamefont {Onnela}},\ }\href@noop {}
	{\bibfield  {journal} {\bibinfo  {journal} {Science}\ }\textbf {\bibinfo
			{volume} {328}},\ \bibinfo {pages} {876} (\bibinfo {year}
		{2010})}\BibitemShut {NoStop}%
	\bibitem [{\citenamefont {Tang}\ \emph {et~al.}(2019)\citenamefont {Tang},
		\citenamefont {Wu}, \citenamefont {L{\"u}}, \citenamefont {Lu},\ and\
		\citenamefont {D'Souza}}]{TAN19}%
	\BibitemOpen
	\bibfield  {author} {\bibinfo {author} {\bibfnamefont {L.}~\bibnamefont
			{Tang}}, \bibinfo {author} {\bibfnamefont {X.}~\bibnamefont {Wu}}, \bibinfo
		{author} {\bibfnamefont {J.}~\bibnamefont {L{\"u}}}, \bibinfo {author}
		{\bibfnamefont {J.}~\bibnamefont {Lu}}, \ and\ \bibinfo {author}
		{\bibfnamefont {R.~M.}\ \bibnamefont {D'Souza}},\ }\href@noop {} {\bibfield
		{journal} {\bibinfo  {journal} {Phys. Rev. E}\ }\textbf {\bibinfo {volume}
			{99}} (\bibinfo {year} {2019})},\ \bibinfo {note} {012304}\BibitemShut
	{NoStop}%
	\bibitem [{\citenamefont {Berner}\ \emph {et~al.}(2020)\citenamefont {Berner},
		\citenamefont {Sawicki},\ and\ \citenamefont {Sch{\"o}ll}}]{BER19b}%
	\BibitemOpen
	\bibfield  {author} {\bibinfo {author} {\bibfnamefont {R.}~\bibnamefont
			{Berner}}, \bibinfo {author} {\bibfnamefont {J.}~\bibnamefont {Sawicki}}, \
		and\ \bibinfo {author} {\bibfnamefont {E.}~\bibnamefont {Sch{\"o}ll}},\
	}\href {\doibase 10.1103/physrevlett.124.088301} {\bibfield  {journal}
		{\bibinfo  {journal} {Phys. Rev. Lett.}\ }\textbf {\bibinfo {volume} {124}},\
		\bibinfo {pages} {088301} (\bibinfo {year} {2020})}\BibitemShut {NoStop}%
	\bibitem [{\citenamefont {Zhang}\ \emph {et~al.}(2015)\citenamefont {Zhang},
		\citenamefont {Boccaletti}, \citenamefont {Guan},\ and\ \citenamefont
		{Liu}}]{ZHA15a}%
	\BibitemOpen
	\bibfield  {author} {\bibinfo {author} {\bibfnamefont {X.}~\bibnamefont
			{Zhang}}, \bibinfo {author} {\bibfnamefont {S.}~\bibnamefont {Boccaletti}},
		\bibinfo {author} {\bibfnamefont {S.}~\bibnamefont {Guan}}, \ and\ \bibinfo
		{author} {\bibfnamefont {Z.}~\bibnamefont {Liu}},\ }\href {\doibase
		10.1103/physrevlett.114.038701} {\bibfield  {journal} {\bibinfo  {journal}
			{Phys. Rev. Lett.}\ }\textbf {\bibinfo {volume} {114}},\ \bibinfo {pages}
		{038701} (\bibinfo {year} {2015})}\BibitemShut {NoStop}%
	\bibitem [{\citenamefont {Nicosia}\ \emph {et~al.}(2013)\citenamefont
		{Nicosia}, \citenamefont {Valencia}, \citenamefont {Chavez}, \citenamefont
		{D{\'i}az-Guilera},\ and\ \citenamefont {Latora}}]{NIC13}%
	\BibitemOpen
	\bibfield  {author} {\bibinfo {author} {\bibfnamefont {V.}~\bibnamefont
			{Nicosia}}, \bibinfo {author} {\bibfnamefont {M.}~\bibnamefont {Valencia}},
		\bibinfo {author} {\bibfnamefont {M.}~\bibnamefont {Chavez}}, \bibinfo
		{author} {\bibfnamefont {A.}~\bibnamefont {D{\'i}az-Guilera}}, \ and\
		\bibinfo {author} {\bibfnamefont {V.}~\bibnamefont {Latora}},\ }\href
	{\doibase 10.1103/physrevlett.110.174102} {\bibfield  {journal} {\bibinfo
			{journal} {Phys. Rev. Lett.}\ }\textbf {\bibinfo {volume} {110}},\ \bibinfo
		{pages} {174102} (\bibinfo {year} {2013})}\BibitemShut {NoStop}%
	\bibitem [{\citenamefont {Gambuzza}\ \emph {et~al.}(2013)\citenamefont
		{Gambuzza}, \citenamefont {Cardillo}, \citenamefont {Fiasconaro},
		\citenamefont {Fortuna}, \citenamefont {G\'omez-Garde\~nes},\ and\
		\citenamefont {Frasca}}]{GAM13}%
	\BibitemOpen
	\bibfield  {author} {\bibinfo {author} {\bibfnamefont {L.~V.}\ \bibnamefont
			{Gambuzza}}, \bibinfo {author} {\bibfnamefont {A.}~\bibnamefont {Cardillo}},
		\bibinfo {author} {\bibfnamefont {A.}~\bibnamefont {Fiasconaro}}, \bibinfo
		{author} {\bibfnamefont {L.}~\bibnamefont {Fortuna}}, \bibinfo {author}
		{\bibfnamefont {J.}~\bibnamefont {G\'omez-Garde\~nes}}, \ and\ \bibinfo
		{author} {\bibfnamefont {M.}~\bibnamefont {Frasca}},\ }\href@noop {}
	{\bibfield  {journal} {\bibinfo  {journal} {Chaos}\ }\textbf {\bibinfo
			{volume} {23}},\ \bibinfo {pages} {043103} (\bibinfo {year}
		{2013})}\BibitemShut {NoStop}%
	\bibitem [{\citenamefont {Zhang}\ \emph {et~al.}(2017)\citenamefont {Zhang},
		\citenamefont {Motter},\ and\ \citenamefont {Nishikawa}}]{ZHA17}%
	\BibitemOpen
	\bibfield  {author} {\bibinfo {author} {\bibfnamefont {L.}~\bibnamefont
			{Zhang}}, \bibinfo {author} {\bibfnamefont {A.~E.}\ \bibnamefont {Motter}}, \
		and\ \bibinfo {author} {\bibfnamefont {T.}~\bibnamefont {Nishikawa}},\ }\href
	{\doibase 10.1103/physrevlett.118.174102} {\bibfield  {journal} {\bibinfo
			{journal} {Phys. Rev. Lett.}\ }\textbf {\bibinfo {volume} {118}},\ \bibinfo
		{pages} {174102} (\bibinfo {year} {2017})}\BibitemShut {NoStop}%
	\bibitem [{\citenamefont {Sawicki}\ \emph {et~al.}(2019)\citenamefont
		{Sawicki}, \citenamefont {Ghosh}, \citenamefont {Jalan},\ and\ \citenamefont
		{Zakharova}}]{SAW19a}%
	\BibitemOpen
	\bibfield  {author} {\bibinfo {author} {\bibfnamefont {J.}~\bibnamefont
			{Sawicki}}, \bibinfo {author} {\bibfnamefont {S.}~\bibnamefont {Ghosh}},
		\bibinfo {author} {\bibfnamefont {S.}~\bibnamefont {Jalan}}, \ and\ \bibinfo
		{author} {\bibfnamefont {A.}~\bibnamefont {Zakharova}},\ }\href@noop {}
	{\bibfield  {journal} {\bibinfo  {journal} {Front. Appl. Math. Stat.}\
		}\textbf {\bibinfo {volume} {5}},\ \bibinfo {pages} {19} (\bibinfo {year}
		{2019})}\BibitemShut {NoStop}%
	\bibitem [{\citenamefont {Winkler}\ \emph {et~al.}(2019)\citenamefont
		{Winkler}, \citenamefont {Sawicki}, \citenamefont {Omelchenko}, \citenamefont
		{Zakharova}, \citenamefont {Anishchenko},\ and\ \citenamefont
		{Sch{\"o}ll}}]{WIN19}%
	\BibitemOpen
	\bibfield  {author} {\bibinfo {author} {\bibfnamefont {M.}~\bibnamefont
			{Winkler}}, \bibinfo {author} {\bibfnamefont {J.}~\bibnamefont {Sawicki}},
		\bibinfo {author} {\bibfnamefont {I.}~\bibnamefont {Omelchenko}}, \bibinfo
		{author} {\bibfnamefont {A.}~\bibnamefont {Zakharova}}, \bibinfo {author}
		{\bibfnamefont {V.}~\bibnamefont {Anishchenko}}, \ and\ \bibinfo {author}
		{\bibfnamefont {E.}~\bibnamefont {Sch{\"o}ll}},\ }\href@noop {} {\bibfield
		{journal} {\bibinfo  {journal} {EPL}\ }\textbf {\bibinfo {volume} {126}},\
		\bibinfo {pages} {50004} (\bibinfo {year} {2019})}\BibitemShut {NoStop}%
	\bibitem [{\citenamefont {Guillery}\ and\ \citenamefont
		{Sherman}(2002)}]{GUI02}%
	\BibitemOpen
	\bibfield  {author} {\bibinfo {author} {\bibfnamefont {R.~W.}\ \bibnamefont
			{Guillery}}\ and\ \bibinfo {author} {\bibfnamefont {S.~M.}\ \bibnamefont
			{Sherman}},\ }\href@noop {} {\bibfield  {journal} {\bibinfo  {journal}
			{Neuron}\ }\textbf {\bibinfo {volume} {33}},\ \bibinfo {pages} {163}
		(\bibinfo {year} {2002})}\BibitemShut {NoStop}%
	\bibitem [{\citenamefont {Bassett}\ \emph {et~al.}(2006)\citenamefont
		{Bassett}, \citenamefont {Meyer-Lindenberg}, \citenamefont {Achard},
		\citenamefont {Duke},\ and\ \citenamefont {Bullmore}}]{BAS06}%
	\BibitemOpen
	\bibfield  {author} {\bibinfo {author} {\bibfnamefont {D.~S.}\ \bibnamefont
			{Bassett}}, \bibinfo {author} {\bibfnamefont {A.}~\bibnamefont
			{Meyer-Lindenberg}}, \bibinfo {author} {\bibfnamefont {S.}~\bibnamefont
			{Achard}}, \bibinfo {author} {\bibfnamefont {T.}~\bibnamefont {Duke}}, \ and\
		\bibinfo {author} {\bibfnamefont {E.~T.}\ \bibnamefont {Bullmore}},\
	}\href@noop {} {\bibfield  {journal} {\bibinfo  {journal} {Proc. Natl. Acad.
				Sci. U.S.A.}\ }\textbf {\bibinfo {volume} {103}},\ \bibinfo {pages} {19518}
		(\bibinfo {year} {2006})}\BibitemShut {NoStop}%
	\bibitem [{\citenamefont {Bassett}\ and\ \citenamefont
		{Bullmore}(2006)}]{BAS06a}%
	\BibitemOpen
	\bibfield  {author} {\bibinfo {author} {\bibfnamefont {D.~S.}\ \bibnamefont
			{Bassett}}\ and\ \bibinfo {author} {\bibfnamefont {E.~T.}\ \bibnamefont
			{Bullmore}},\ }\href {\doibase 10.1177/1073858406293182} {\bibfield
		{journal} {\bibinfo  {journal} {Neuroscientist}\ }\textbf {\bibinfo {volume}
			{12}},\ \bibinfo {pages} {512} (\bibinfo {year} {2006})}\BibitemShut
	{NoStop}%
	\bibitem [{\citenamefont {Gambuzza}\ \emph {et~al.}(2016)\citenamefont
		{Gambuzza}, \citenamefont {Frasca}, \citenamefont {Fortuna},\ and\
		\citenamefont {Boccaletti}}]{GAM16b}%
	\BibitemOpen
	\bibfield  {author} {\bibinfo {author} {\bibfnamefont {L.~V.}\ \bibnamefont
			{Gambuzza}}, \bibinfo {author} {\bibfnamefont {M.}~\bibnamefont {Frasca}},
		\bibinfo {author} {\bibfnamefont {L.}~\bibnamefont {Fortuna}}, \ and\
		\bibinfo {author} {\bibfnamefont {S.}~\bibnamefont {Boccaletti}},\
	}\href@noop {} {\bibfield  {journal} {\bibinfo  {journal} {Phys. Rev. E}\
		}\textbf {\bibinfo {volume} {93}},\ \bibinfo {pages} {042203} (\bibinfo
		{year} {2016})}\BibitemShut {NoStop}%
	\bibitem [{\citenamefont {Pecora}\ \emph {et~al.}(2014)\citenamefont {Pecora},
		\citenamefont {Sorrentino}, \citenamefont {Hagerstrom}, \citenamefont
		{Murphy},\ and\ \citenamefont {Roy}}]{PEC14}%
	\BibitemOpen
	\bibfield  {author} {\bibinfo {author} {\bibfnamefont {L.~M.}\ \bibnamefont
			{Pecora}}, \bibinfo {author} {\bibfnamefont {F.}~\bibnamefont {Sorrentino}},
		\bibinfo {author} {\bibfnamefont {A.~M.}\ \bibnamefont {Hagerstrom}},
		\bibinfo {author} {\bibfnamefont {T.~E.}\ \bibnamefont {Murphy}}, \ and\
		\bibinfo {author} {\bibfnamefont {R.}~\bibnamefont {Roy}},\ }\href@noop {}
	{\bibfield  {journal} {\bibinfo  {journal} {Nat. Commun.}\ }\textbf {\bibinfo
			{volume} {5}},\ \bibinfo {pages} {4079} (\bibinfo {year} {2014})}\BibitemShut
	{NoStop}%
	\bibitem [{\citenamefont {Kuramoto}\ and\ \citenamefont
		{Battogtokh}(2002)}]{KUR02a}%
	\BibitemOpen
	\bibfield  {author} {\bibinfo {author} {\bibfnamefont {Y.}~\bibnamefont
			{Kuramoto}}\ and\ \bibinfo {author} {\bibfnamefont {D.}~\bibnamefont
			{Battogtokh}},\ }\href@noop {} {\bibfield  {journal} {\bibinfo  {journal}
			{Nonlin. Phen. in Complex Sys.}\ }\textbf {\bibinfo {volume} {5}},\ \bibinfo
		{pages} {380} (\bibinfo {year} {2002})}\BibitemShut {NoStop}%
	\bibitem [{\citenamefont {Abrams}\ and\ \citenamefont
		{Strogatz}(2004)}]{ABR04}%
	\BibitemOpen
	\bibfield  {author} {\bibinfo {author} {\bibfnamefont {D.~M.}\ \bibnamefont
			{Abrams}}\ and\ \bibinfo {author} {\bibfnamefont {S.~H.}\ \bibnamefont
			{Strogatz}},\ }\href {\doibase 10.1103/physrevlett.93.174102} {\bibfield
		{journal} {\bibinfo  {journal} {Phys. Rev. Lett.}\ }\textbf {\bibinfo
			{volume} {93}},\ \bibinfo {pages} {174102} (\bibinfo {year}
		{2004})}\BibitemShut {NoStop}%
	\bibitem [{\citenamefont {Panaggio}\ and\ \citenamefont
		{Abrams}(2015)}]{PAN15}%
	\BibitemOpen
	\bibfield  {author} {\bibinfo {author} {\bibfnamefont {M.~J.}\ \bibnamefont
			{Panaggio}}\ and\ \bibinfo {author} {\bibfnamefont {D.~M.}\ \bibnamefont
			{Abrams}},\ }\href {\doibase 10.1088/0951-7715/28/3/r67} {\bibfield
		{journal} {\bibinfo  {journal} {Nonlinearity}\ }\textbf {\bibinfo {volume}
			{28}},\ \bibinfo {pages} {R67} (\bibinfo {year} {2015})}\BibitemShut
	{NoStop}%
	\bibitem [{\citenamefont {Sch{\"o}ll}(2016)}]{SCH16b}%
	\BibitemOpen
	\bibfield  {author} {\bibinfo {author} {\bibfnamefont {E.}~\bibnamefont
			{Sch{\"o}ll}},\ }\href {\doibase 10.1140/epjst/e2016-02646-3} {\bibfield
		{journal} {\bibinfo  {journal} {Eur. Phys. J. Spec. Top.}\ }\textbf {\bibinfo
			{volume} {225}},\ \bibinfo {pages} {891} (\bibinfo {year}
		{2016})}\BibitemShut {NoStop}%
	\bibitem [{\citenamefont {Omel'chenko}(2018)}]{OME18a}%
	\BibitemOpen
	\bibfield  {author} {\bibinfo {author} {\bibfnamefont {O.~E.}\ \bibnamefont
			{Omel'chenko}},\ }\href {\doibase 10.1088/1261-6544/aaaa07} {\bibfield
		{journal} {\bibinfo  {journal} {Nonlinearity}\ }\textbf {\bibinfo {volume}
			{31}},\ \bibinfo {pages} {R121} (\bibinfo {year} {2018})}\BibitemShut
	{NoStop}%
	\bibitem [{\citenamefont {Omel'chenko}\ and\ \citenamefont
		{Knobloch}(2019)}]{OME19c}%
	\BibitemOpen
	\bibfield  {author} {\bibinfo {author} {\bibfnamefont {O.~E.}\ \bibnamefont
			{Omel'chenko}}\ and\ \bibinfo {author} {\bibfnamefont {E.}~\bibnamefont
			{Knobloch}},\ }\href {\doibase 10.1088/1367-2630/ab3f6b} {\bibfield
		{journal} {\bibinfo  {journal} {New J. Phys.}\ }\textbf {\bibinfo {volume}
			{21}},\ \bibinfo {pages} {093034} (\bibinfo {year} {2019})}\BibitemShut
	{NoStop}%
	\bibitem [{\citenamefont {Sch{\"o}ll}\ \emph {et~al.}(2020)\citenamefont
		{Sch{\"o}ll}, \citenamefont {Zakharova},\ and\ \citenamefont
		{Andrzejak}}]{SCH20b}%
	\BibitemOpen
	\bibfield  {author} {\bibinfo {author} {\bibfnamefont {E.}~\bibnamefont
			{Sch{\"o}ll}}, \bibinfo {author} {\bibfnamefont {A.}~\bibnamefont
			{Zakharova}}, \ and\ \bibinfo {author} {\bibfnamefont {R.~G.}\ \bibnamefont
			{Andrzejak}},\ }\href {\doibase 10.3389/978-2-88963-311-1} {\emph {\bibinfo
			{title} {Chimera States in Complex Networks}}},\ Research Topics, Front.
	Appl. Math. Stat.\ (\bibinfo  {publisher} {Lausanne: Frontiers Media SA},\
	\bibinfo {year} {2020})\ \bibinfo {note} {ebook}\BibitemShut {NoStop}%
	\bibitem [{\citenamefont {Rattenborg}\ \emph {et~al.}(2000)\citenamefont
		{Rattenborg}, \citenamefont {Amlaner},\ and\ \citenamefont {Lima}}]{RAT00}%
	\BibitemOpen
	\bibfield  {author} {\bibinfo {author} {\bibfnamefont {N.~C.}\ \bibnamefont
			{Rattenborg}}, \bibinfo {author} {\bibfnamefont {C.~J.}\ \bibnamefont
			{Amlaner}}, \ and\ \bibinfo {author} {\bibfnamefont {S.~L.}\ \bibnamefont
			{Lima}},\ }\href@noop {} {\bibfield  {journal} {\bibinfo  {journal}
			{Neurosci. Biobehav. Rev.}\ }\textbf {\bibinfo {volume} {24}},\ \bibinfo
		{pages} {817} (\bibinfo {year} {2000})}\BibitemShut {NoStop}%
	\bibitem [{\citenamefont {Rattenborg}\ \emph {et~al.}(2016)\citenamefont
		{Rattenborg}, \citenamefont {Voirin}, \citenamefont {Cruz}, \citenamefont
		{Tisdale}, \citenamefont {Dell'Omo}, \citenamefont {Lipp}, \citenamefont
		{Wikelski},\ and\ \citenamefont {Vyssotski}}]{RAT16}%
	\BibitemOpen
	\bibfield  {author} {\bibinfo {author} {\bibfnamefont {N.~C.}\ \bibnamefont
			{Rattenborg}}, \bibinfo {author} {\bibfnamefont {B.}~\bibnamefont {Voirin}},
		\bibinfo {author} {\bibfnamefont {S.~M.}\ \bibnamefont {Cruz}}, \bibinfo
		{author} {\bibfnamefont {R.}~\bibnamefont {Tisdale}}, \bibinfo {author}
		{\bibfnamefont {G.}~\bibnamefont {Dell'Omo}}, \bibinfo {author}
		{\bibfnamefont {H.~P.}\ \bibnamefont {Lipp}}, \bibinfo {author}
		{\bibfnamefont {M.}~\bibnamefont {Wikelski}}, \ and\ \bibinfo {author}
		{\bibfnamefont {A.~L.}\ \bibnamefont {Vyssotski}},\ }\href {\doibase
		10.1038/ncomms12468} {\bibfield  {journal} {\bibinfo  {journal} {Nat.
				Commun.}\ }\textbf {\bibinfo {volume} {7}},\ \bibinfo {pages} {12468}
		(\bibinfo {year} {2016})}\BibitemShut {NoStop}%
	\bibitem [{\citenamefont {Mascetti}(2016)}]{MAS16}%
	\BibitemOpen
	\bibfield  {author} {\bibinfo {author} {\bibfnamefont {G.~G.}\ \bibnamefont
			{Mascetti}},\ }\href@noop {} {\bibfield  {journal} {\bibinfo  {journal} {Nat
				Sci Sleep}\ }\textbf {\bibinfo {volume} {8}},\ \bibinfo {pages} {221}
		(\bibinfo {year} {2016})}\BibitemShut {NoStop}%
	\bibitem [{\citenamefont {Nikolaev}\ \emph {et~al.}(2010)\citenamefont
		{Nikolaev}, \citenamefont {Gepshtein}, \citenamefont {Gong},\ and\
		\citenamefont {van Leeuwen}}]{NIK10a}%
	\BibitemOpen
	\bibfield  {author} {\bibinfo {author} {\bibfnamefont {A.~R.}\ \bibnamefont
			{Nikolaev}}, \bibinfo {author} {\bibfnamefont {S.}~\bibnamefont {Gepshtein}},
		\bibinfo {author} {\bibfnamefont {P.}~\bibnamefont {Gong}}, \ and\ \bibinfo
		{author} {\bibfnamefont {C.}~\bibnamefont {van Leeuwen}},\ }\href@noop {}
	{\bibfield  {journal} {\bibinfo  {journal} {Cerebral Cortex}\ }\textbf
		{\bibinfo {volume} {20}},\ \bibinfo {pages} {365} (\bibinfo {year}
		{2010})}\BibitemShut {NoStop}%
	\bibitem [{\citenamefont {Ahn}\ and\ \citenamefont {Rubchinsky}(2013)}]{AHN13}%
	\BibitemOpen
	\bibfield  {author} {\bibinfo {author} {\bibfnamefont {S.}~\bibnamefont
			{Ahn}}\ and\ \bibinfo {author} {\bibfnamefont {L.~L.}\ \bibnamefont
			{Rubchinsky}},\ }\href@noop {} {\bibfield  {journal} {\bibinfo  {journal}
			{Chaos}\ }\textbf {\bibinfo {volume} {23}},\ \bibinfo {pages} {013138}
		(\bibinfo {year} {2013})}\BibitemShut {NoStop}%
	\bibitem [{\citenamefont {Ahn}\ \emph {et~al.}(2014)\citenamefont {Ahn},
		\citenamefont {Rubchinsky},\ and\ \citenamefont {Lapish}}]{AHN14}%
	\BibitemOpen
	\bibfield  {author} {\bibinfo {author} {\bibfnamefont {S.}~\bibnamefont
			{Ahn}}, \bibinfo {author} {\bibfnamefont {L.~L.}\ \bibnamefont {Rubchinsky}},
		\ and\ \bibinfo {author} {\bibfnamefont {C.~C.}\ \bibnamefont {Lapish}},\
	}\href@noop {} {\bibfield  {journal} {\bibinfo  {journal} {Cerebral Cortex}\
		}\textbf {\bibinfo {volume} {24}},\ \bibinfo {pages} {2553} (\bibinfo {year}
		{2014})}\BibitemShut {NoStop}%
	\bibitem [{\citenamefont {Jiruska}\ \emph {et~al.}(2013)\citenamefont
		{Jiruska}, \citenamefont {de~Curtis}, \citenamefont {Jefferys}, \citenamefont
		{Schevon}, \citenamefont {Schiff},\ and\ \citenamefont {Schindler}}]{JIR13}%
	\BibitemOpen
	\bibfield  {author} {\bibinfo {author} {\bibfnamefont {P.}~\bibnamefont
			{Jiruska}}, \bibinfo {author} {\bibfnamefont {M.}~\bibnamefont {de~Curtis}},
		\bibinfo {author} {\bibfnamefont {J.~G.~R.}\ \bibnamefont {Jefferys}},
		\bibinfo {author} {\bibfnamefont {C.~A.}\ \bibnamefont {Schevon}}, \bibinfo
		{author} {\bibfnamefont {S.~J.}\ \bibnamefont {Schiff}}, \ and\ \bibinfo
		{author} {\bibfnamefont {K.}~\bibnamefont {Schindler}},\ }\href@noop {}
	{\bibfield  {journal} {\bibinfo  {journal} {J. Physiol.}\ }\textbf {\bibinfo
			{volume} {591.4}},\ \bibinfo {pages} {787} (\bibinfo {year}
		{2013})}\BibitemShut {NoStop}%
	\bibitem [{\citenamefont {Jirsa}\ \emph {et~al.}(2014)\citenamefont {Jirsa},
		\citenamefont {Stacey}, \citenamefont {Quilichini}, \citenamefont {Ivanov},\
		and\ \citenamefont {Bernard}}]{JIR14}%
	\BibitemOpen
	\bibfield  {author} {\bibinfo {author} {\bibfnamefont {V.~K.}\ \bibnamefont
			{Jirsa}}, \bibinfo {author} {\bibfnamefont {W.~C.}\ \bibnamefont {Stacey}},
		\bibinfo {author} {\bibfnamefont {P.~P.}\ \bibnamefont {Quilichini}},
		\bibinfo {author} {\bibfnamefont {A.~I.}\ \bibnamefont {Ivanov}}, \ and\
		\bibinfo {author} {\bibfnamefont {C.}~\bibnamefont {Bernard}},\ }\href@noop
	{} {\bibfield  {journal} {\bibinfo  {journal} {Brain}\ }\textbf {\bibinfo
			{volume} {137}},\ \bibinfo {pages} {2210} (\bibinfo {year}
		{2014})}\BibitemShut {NoStop}%
	\bibitem [{\citenamefont {Sakaguchi}(2006)}]{SAK06a}%
	\BibitemOpen
	\bibfield  {author} {\bibinfo {author} {\bibfnamefont {H.}~\bibnamefont
			{Sakaguchi}},\ }\href {\doibase 10.1103/physreve.73.031907} {\bibfield
		{journal} {\bibinfo  {journal} {Phys. Rev. E}\ }\textbf {\bibinfo {volume}
			{73}},\ \bibinfo {eid} {031907} (\bibinfo {year} {2006})}\BibitemShut
	{NoStop}%
	\bibitem [{\citenamefont {Omelchenko}\ \emph {et~al.}(2013)\citenamefont
		{Omelchenko}, \citenamefont {Omel'chenko}, \citenamefont {H\"{o}vel},\ and\
		\citenamefont {Sch{\"o}ll}}]{OME13}%
	\BibitemOpen
	\bibfield  {author} {\bibinfo {author} {\bibfnamefont {I.}~\bibnamefont
			{Omelchenko}}, \bibinfo {author} {\bibfnamefont {O.~E.}\ \bibnamefont
			{Omel'chenko}}, \bibinfo {author} {\bibfnamefont {P.}~\bibnamefont
			{H\"{o}vel}}, \ and\ \bibinfo {author} {\bibfnamefont {E.}~\bibnamefont
			{Sch{\"o}ll}},\ }\href {\doibase 10.1103/physrevlett.110.224101} {\bibfield
		{journal} {\bibinfo  {journal} {Phys. Rev. Lett.}\ }\textbf {\bibinfo
			{volume} {110}},\ \bibinfo {pages} {224101} (\bibinfo {year}
		{2013})}\BibitemShut {NoStop}%
	\bibitem [{\citenamefont {Hizanidis}\ \emph {et~al.}(2014)\citenamefont
		{Hizanidis}, \citenamefont {Kanas}, \citenamefont {Bezerianos},\ and\
		\citenamefont {Bountis}}]{HIZ13}%
	\BibitemOpen
	\bibfield  {author} {\bibinfo {author} {\bibfnamefont {J.}~\bibnamefont
			{Hizanidis}}, \bibinfo {author} {\bibfnamefont {V.}~\bibnamefont {Kanas}},
		\bibinfo {author} {\bibfnamefont {A.}~\bibnamefont {Bezerianos}}, \ and\
		\bibinfo {author} {\bibfnamefont {T.}~\bibnamefont {Bountis}},\ }\href
	{\doibase 10.1142/s0218127414500308} {\bibfield  {journal} {\bibinfo
			{journal} {Int. J. Bifurcation Chaos}\ }\textbf {\bibinfo {volume} {24}},\
		\bibinfo {pages} {1450030} (\bibinfo {year} {2014})}\BibitemShut {NoStop}%
	\bibitem [{\citenamefont {Omelchenko}\ \emph {et~al.}(2015)\citenamefont
		{Omelchenko}, \citenamefont {Provata}, \citenamefont {Hizanidis},
		\citenamefont {Sch{\"o}ll},\ and\ \citenamefont {H\"{o}vel}}]{OME15}%
	\BibitemOpen
	\bibfield  {author} {\bibinfo {author} {\bibfnamefont {I.}~\bibnamefont
			{Omelchenko}}, \bibinfo {author} {\bibfnamefont {A.}~\bibnamefont {Provata}},
		\bibinfo {author} {\bibfnamefont {J.}~\bibnamefont {Hizanidis}}, \bibinfo
		{author} {\bibfnamefont {E.}~\bibnamefont {Sch{\"o}ll}}, \ and\ \bibinfo
		{author} {\bibfnamefont {P.}~\bibnamefont {H\"{o}vel}},\ }\href {\doibase
		10.1103/physreve.91.022917} {\bibfield  {journal} {\bibinfo  {journal} {Phys.
				Rev. E}\ }\textbf {\bibinfo {volume} {91}},\ \bibinfo {pages} {022917}
		(\bibinfo {year} {2015})}\BibitemShut {NoStop}%
	\bibitem [{\citenamefont {Hizanidis}\ \emph {et~al.}(2016)\citenamefont
		{Hizanidis}, \citenamefont {Kouvaris}, \citenamefont {Zamora-L{\'o}pez},
		\citenamefont {D{\'i}az-Guilera},\ and\ \citenamefont
		{Antonopoulos}}]{HIZ16}%
	\BibitemOpen
	\bibfield  {author} {\bibinfo {author} {\bibfnamefont {J.}~\bibnamefont
			{Hizanidis}}, \bibinfo {author} {\bibfnamefont {N.~E.}\ \bibnamefont
			{Kouvaris}}, \bibinfo {author} {\bibfnamefont {G.}~\bibnamefont
			{Zamora-L{\'o}pez}}, \bibinfo {author} {\bibfnamefont {A.}~\bibnamefont
			{D{\'i}az-Guilera}}, \ and\ \bibinfo {author} {\bibfnamefont
			{C.}~\bibnamefont {Antonopoulos}},\ }\href {\doibase 10.1038/srep19845}
	{\bibfield  {journal} {\bibinfo  {journal} {Sci. Rep.}\ }\textbf {\bibinfo
			{volume} {6}},\ \bibinfo {pages} {19845} (\bibinfo {year}
		{2016})}\BibitemShut {NoStop}%
	\bibitem [{\citenamefont {FitzHugh}(1961)}]{FIT61}%
	\BibitemOpen
	\bibfield  {author} {\bibinfo {author} {\bibfnamefont {R.}~\bibnamefont
			{FitzHugh}},\ }\href@noop {} {\bibfield  {journal} {\bibinfo  {journal}
			{Biophys. J.}\ }\textbf {\bibinfo {volume} {1}},\ \bibinfo {pages} {445}
		(\bibinfo {year} {1961})}\BibitemShut {NoStop}%
	\bibitem [{\citenamefont {Watts}\ and\ \citenamefont {Strogatz}(1998)}]{WAT98}%
	\BibitemOpen
	\bibfield  {author} {\bibinfo {author} {\bibfnamefont {D.~J.}\ \bibnamefont
			{Watts}}\ and\ \bibinfo {author} {\bibfnamefont {S.~H.}\ \bibnamefont
			{Strogatz}},\ }\href@noop {} {\bibfield  {journal} {\bibinfo  {journal}
			{Nature}\ }\textbf {\bibinfo {volume} {393}},\ \bibinfo {pages} {440}
		(\bibinfo {year} {1998})}\BibitemShut {NoStop}%
	\bibitem [{\citenamefont {Arenas}\ \emph {et~al.}(2008)\citenamefont {Arenas},
		\citenamefont {D{\'i}az-Guilera}, \citenamefont {Kurths}, \citenamefont
		{Moreno},\ and\ \citenamefont {Zhou}}]{ARE08}%
	\BibitemOpen
	\bibfield  {author} {\bibinfo {author} {\bibfnamefont {A.}~\bibnamefont
			{Arenas}}, \bibinfo {author} {\bibfnamefont {A.}~\bibnamefont
			{D{\'i}az-Guilera}}, \bibinfo {author} {\bibfnamefont {J.}~\bibnamefont
			{Kurths}}, \bibinfo {author} {\bibfnamefont {Y.}~\bibnamefont {Moreno}}, \
		and\ \bibinfo {author} {\bibfnamefont {C.}~\bibnamefont {Zhou}},\ }\href
	{\doibase doi: 10.1016/j.physrep.2008.09.002} {\bibfield  {journal} {\bibinfo
			{journal} {Phys. Rep.}\ }\textbf {\bibinfo {volume} {469}},\ \bibinfo
		{pages} {93} (\bibinfo {year} {2008})}\BibitemShut {NoStop}%
	\bibitem [{\citenamefont {Barahona}\ and\ \citenamefont
		{Pecora}(2002)}]{BAR02}%
	\BibitemOpen
	\bibfield  {author} {\bibinfo {author} {\bibfnamefont {M.}~\bibnamefont
			{Barahona}}\ and\ \bibinfo {author} {\bibfnamefont {L.~M.}\ \bibnamefont
			{Pecora}},\ }\href {\doibase 10.1103/physrevlett.89.054101} {\bibfield
		{journal} {\bibinfo  {journal} {Phys. Rev. Lett.}\ }\textbf {\bibinfo
			{volume} {89}},\ \bibinfo {pages} {054101} (\bibinfo {year}
		{2002})}\BibitemShut {NoStop}%
	\bibitem [{\citenamefont {Hong}\ \emph {et~al.}(2004)\citenamefont {Hong},
		\citenamefont {Jun~Kim}, \citenamefont {Choi},\ and\ \citenamefont
		{Park}}]{HON04}%
	\BibitemOpen
	\bibfield  {author} {\bibinfo {author} {\bibfnamefont {H.}~\bibnamefont
			{Hong}}, \bibinfo {author} {\bibfnamefont {B.}~\bibnamefont {Jun~Kim}},
		\bibinfo {author} {\bibfnamefont {M.~Y.}\ \bibnamefont {Choi}}, \ and\
		\bibinfo {author} {\bibfnamefont {H.}~\bibnamefont {Park}},\ }\href@noop {}
	{\bibfield  {journal} {\bibinfo  {journal} {Phys. Rev. E}\ }\textbf {\bibinfo
			{volume} {69}},\ \bibinfo {pages} {067105} (\bibinfo {year}
		{2004})}\BibitemShut {NoStop}%
\end{thebibliography}
\end{document}